\documentclass[preprint,12pt,a4paper,openright]{elsarticle}

\usepackage{setspace} 

\makeatletter
\def\ps@pprintTitle{%
	\let\@oddhead\@empty
	\let\@evenhead\@empty
	\def\@oddfoot{\reset@font\hfil\thepage\hfil}
	\let\@evenfoot\@oddfoot
}
\makeatother

\usepackage{mathptmx}	
\usepackage[utf8]{inputenc}
\usepackage[english]{babel}
\usepackage{amsthm,amssymb,amsmath,amsbsy}
\usepackage{bigints}
\usepackage{multirow}
\usepackage{xcolor}
\usepackage{mathrsfs}
\usepackage{graphicx}
\usepackage{epsfig}
\usepackage{caption}
\usepackage{subcaption}
\usepackage{lscape} 

\usepackage{float}

\usepackage{booktabs,caption}
\usepackage[flushleft]{threeparttable}

\usepackage{chngcntr} 
\counterwithin{table}{section} 
\counterwithin{figure}{section} 

\usepackage{natbib} 
\biboptions{comma,round,authoryear}
\usepackage[colorlinks=true]{hyperref}
\hypersetup{
	colorlinks=true,
	linkcolor=black,  
	citecolor=black,
}

\newtheorem{thm}{Theorem}[section]
\newtheorem{dfntn}{Definition}[section]
\newtheorem{cor}{Corollary}[section]
\newtheorem{app}{Appendix}[section]

\title{On Modeling Bivariate Left Censored Data using Reversed Hazard Rates.}
\author{{Durga Vasudevan$^{1}$ and G. Asha*}\\
	{\textit{Department of Statistics}}\\
	{\textit{Cochin University of Science and Technology, Cochin, Kerala, India-682022}}\\
	{*Corresponding Author Email: \textit{asha.gopalakrishnan@gmail.com}}\\
	{$^{1}$Email: \textit{durgavvn10@gmail.com}}}

\begin{document}
	
	\begin{frontmatter}
		\begin{abstract}			
			When the observations are not quantified and are known to be less than a threshold value, the concept of left censoring needs to be included in the analysis of such datasets. In many real multi component lifetime  systems left censored data is very common. The usual assumption that components which are part of a system, work independently seems not appropriate in a number of applications. For instance it is more realistic to acknowledge that the working status of a component affects the remaining components. When you have left-censored data, it is more meaningful to use the reversed hazard rate, proposed as a dual to the hazard rate. In this paper, we propose a model for left-censored bivariate data incorporating the  dependence enjoyed among the components, based on a dynamic bivariate vector reversed hazard rate proposed in \cite{gurler1996bivariate}. The properties of the proposed model is studied. The maximum likelihood method of estimation is shown to work well for moderately large samples. The Bayesian approach to the estimation of parameters is also presented. The complexity of the likelihood function is handled through the Metropolis - Hastings algorithm. This is executed with the MH adaptive package in r. Different interval estimation techniques of the parameters are also considered. Applications of this model is demonstrated by illustrating the usefulness of the model in analyzing real data.
		\end{abstract}
	\end{frontmatter}
	
\section{Introduction}
	A lifetime $Y$ associated with a subject is said to be left-censored if the event of interest has already occurred before that individual/unit is considered in the study at time $C_l$. The exact lifetime $Y$ will be known only if $Y \geq C_l$ and hence a left-censored data is represented by a pair of random variables $(Y,\delta)$ where $\delta=1$ if the event is observed and $0$ otherwise. The observed data is $T=\text{max}\{Y,C_l\}$. Left-censored data have immense application in survival/reliability studies. They occur in life-test applications when a unit has failed at the time of its first inspection. They are also very common in bio-monitoring/environmental studies where observations could lie below a threshold value called limit of detection (LOD). 
	\par There are various approaches for analysing  left-censored data. There are many naive approaches like $\beta$-substitution method (\cite{ganser2010accurate}) where each non-detected observation is replaced with $LOD$, $\frac{LOD}{2}$ or $\frac{LOD}{\sqrt{2}}$ and model based imputation techniques (\cite{krishnamoorthy2009model}). In a non-parametric approach, a reverse algorithm of the Kaplan-Meier (\cite{ware1976reanalysis}) helps to handle the left-censored data. The most conventional way to estimate the parameters is using the maximum likelihood estimation technique where the observed data contributes to the likelihood function through the probability density function (pdf) and the left-censored data contributes through the cumulative distribution function (CDF). The likelihood can be represented in terms of the reversed hazard rates too (\cite{lawless2011statistical}). The reversed hazard rate was proposed by \cite{barlow1963properties} as a dual to the hazard rate. The concept of reversed hazard rate has been a subject of extensive study (see \cite{keilson1982uniform, block1998reversed}) since the usefulness of the  reversed hazard rates was reported  in the estimation of distribution function in the presence of left-censored observations (\cite{ware1976reanalysis}).
	\par Let $a=\text{inf}\{y~|F(y)>0\}$ and $b=\text{sup}\{y~|F(y)<1\}$. Then $(a,b)$ where $-\infty \leq a <b \leq \infty$ is the interval of support of $Y$ with the distribution function $F(y)$. The reversed hazard rate of $Y,$ denoted as $r(y),$ is defined for $y>a$ as,
	\begin{equation}\label{rhrdef}
		r(y)= \lim_{\Delta y \to 0} \frac{P(y-\Delta y < Y \leq y|Y \leq y)}{\Delta y}.
	\end{equation}
	Note that $r(y)\Delta y$ is the probability of failure in a small interval $(y-\Delta y, y]$, given that the failure has occurred before $y$. When the probability density function of $Y$ exists, \eqref{rhrdef} can be expressed as,
	\begin{equation*}\label{rhrdef2}
		r(y)=\frac{f(y)}{F(y)}=\frac{d}{dy}log F(y).
	\end{equation*}
	\cite{keilson1982uniform} showed that the reversed hazard rate, $r(y)$, uniquely determines the distribution function through the relation,
	\begin{equation*}
		F(y)=exp\big [-\int_{y}^{b} r(u) du \big ].
	\end{equation*}
	The probability density function, $f(y)$, and the cumulative reversed hazard rate, $R(y)$, can be obtained using the relations,
	\begin{equation*}
		f(y)=r(y) exp\big [-\int_{y}^{b} r(u) du \big ]~\text{and}~R(y)=\int_{y}^{b} r(u) du.
	\end{equation*}	
	\par The concept of reversed hazard and its uses has been extensively studied in literature. For some recent works we refer to \cite{hanagal2017modeling, hanagal2019shared, pandey2020analysis} and \cite{hanagal2021correlated}.
	\par This concept was extended to higher dimensions by many authors. \cite{gurler1996bivariate} gave a three-component bivariate reversed hazard vector analogous to the bivariate hazard vector introduced by \cite{dabrowska1988kaplan} which can be used for the estimation of bivariate distribution function, $F(y_1,y_2)$, when the lifetime data is right truncated. \cite{roy2002characterization} and \cite{pg2006bivariate} defined the bivariate reversed hazard rate as a two component vector while a scalar extension of the same is found in \cite{bismi2005bivariate}.
	\par \cite{gupta1998modeling} proposed a dual model called proportional reversed hazards (PRH) model, which is expressed as,
	\begin{equation}\label{prhr}
	r(y)=\theta r_0(y),
	\end{equation}
	where $\theta>0$ and $r_0$ is the baseline reversed hazard rate. The corresponding distribution function is,
	\begin{equation*}
	F(y)=[F_0(y)]^\theta,
	\end{equation*}
	where $F_0(y)$ is the baseline distribution function.
	The model in \eqref{prhr} is helpful in the analysis of left-censored or right-truncated data. The PRH model has some extremely interesting properties. The parameter `$\theta$' is crucial in maintaining the structural properties of the baseline distribution. It is used to manage the skewness of the distribution. Also ageing and relative ageing properties of PRH models were studied extensively in \cite{di2000some}. For recent works on this model, see \cite{balakrishnan2021comparisons} and \cite{popovic2021generalized}. This PRH model has been extended to the bivariate case and studied extensively by \cite{kundu2010class} where their marginals follow univariate proportional hazards model. They defined the joint distribution of $(Y_1,Y_2)$ for $y_1>0$ and $y_2>0$ in this case as,
	\begin{equation}
		F(y_1,y_2)=(F_{0}(y_1))^{\theta_1}(F_{0}(y_2))^{\theta_2}(F_{0}(z))^{\theta_3},
	\end{equation}
	where $z=\text{min}\{y_1,y_2\}$. 
	But when a data exhibits an inherent dependence among the components we extend the notion of PRH to take into consideration this fact. The main aim of this paper is to generalize the Bivariate Proportional Reversed Hazard Model (BPRHM) introduced by \cite{kundu2010class} to incorporate an inherent dependence in the data. In this paper, we consider the load share dependence. Here when one of the variables has undergone an event of interest, the distribution of the other is affected. A typical example could be lifetime of pair of eyes and kidneys. Another example could be same events affecting a twin and co-twin.	
	\par Accordingly in Section \ref{secmodel}, we propose and study a new bivariate model to capture the lifetime behaviour of a two-component system having the load share dependence when the observations obtained are left-censored. The proposed model is a general class of model and enjoys some very good properties making it simple to generate samples from the same. In Section \ref{properties}, various properties including the identifiability of the model is studied. In Section \ref{estimation}, the maximum likelihood estimation and Bayesian estimation methodology of the parameters are discussed. In Section \ref{interval}, we explain the different interval estimation techniques used in the analysis. In Section \ref{simulation}, these methodologies are validated using simulation studies. Accordingly in Section \ref{realdata}, the Australian twin data is analysed. Studies have established that there exists a dependence between a twin and a co-twin on many characteristics (\cite{paluszny1974twin}, \cite{fortuna2010twin}). We establish the load share dependence between the twins through a likelihood ratio test and the suitability of our proposed model. The final Section \ref{conclusion} gives a conclusion and discussion of the work done in this paper.
	
	\section{The Model Construction} \label{secmodel}
	Consider a parallel system with two different but associated components having lifetimes $Y_1$ and $Y_2$. We assume that both the components are simultaneously working and their corresponding lifetimes are independently and identically distributed until one of them fails and the remaining system works with a renewed parameter. The system fails on the failure of both the components. We are interested in the failure of the first component given the time to failure of the system. The metric proposed by \cite{gurler1996bivariate} is modified and the reversed hazard rate is defined as a vector $\pmb{\lambda}(\underline{y})=\big(\lambda_{10}(y)+\lambda_{20}(y),\lambda_{12}(y_1|y_2),\lambda_{21}(y_2|y_1)\big)$ where
	\begin{equation*}
		\lambda_{i,3-i}(y_i|y_{3-i}) = \lim_{\Delta y_{i} \to 0^{+}} \frac{P(y_i-\Delta y_i < Y_i \leq y_i|Y_i \leq y_i, Y_{3-i} = y_{3-i})}{\Delta y_i};~y_i<y_{3-i},~i=1,2,
	\end{equation*}
	where $\lambda_{i,3-i}(y_i|y_{3-i})dy$ denote the probability of the $i^{th}$ component to have failed in the interval $(y_{i}-\Delta y_{i},y_{i})$ given that it has failed by the time $y_{i}$ and the $(3-i)^{th}$ component has failed at $y_{3-i};~i=1,2$. Note that for $y_{i} < y_{3-i}$, $\lambda_{3-i,i}(y_{3-i}|y_{i}) = \lambda_{3-i,0}(y)$ for $i=1,2$ where 
	\begin{equation*}
		\lambda_{i0}(y)= \lim_{\Delta y \to 0^{+}} \frac{P(y-\Delta y < Y_i \leq y|Y_1 \leq y, Y_2 \leq y)}{\Delta y};~y_i=y,~i=1,2
	\end{equation*}
	denote the probability that either of the components has failed in the interval $(y-\Delta y, y)$ given both the components have failed at $y$. The vector $\pmb{\lambda}(\underline{y})$ uniquely determines $f(y_1,y_2)$ through the relation,
	\begin{equation}\label{coxpdf}
		f(y_1,y_2)=\lambda_{10}(y_1) \lambda_{21}(y_2|y_1) exp \Big [- \int_{y_2}^{y_1} \lambda_{21}(u|y_1)du - \int_{y_1}^{b} \{\lambda_{10}(u)+\lambda_{20}(u)\}du \Big ].
	\end{equation}
	for $y_1>y_2$ with an analogous expression for $y_2>y_1$.
	\par Under proportional reversed hazards assumption, the reversed hazard rates are assumed to be proportional to a baseline reversed hazard function, $r_0(\cdot)$.
	\begin{align} \label{prhrd}
	\begin{split}
	\lambda_{10}(y)&=\theta_1 r_0(y);~~~y_1=y_2=y \\
	\lambda_{20}(y)&=\theta_2 r_0(y);~~~y_1=y_2=y \\
	\lambda_{12}(y_1|y_2)&=\theta_1^{\prime} r_0(y_1);~~~y_1<y_2 \\
	\lambda_{21}(y_2|y_1)&=\theta_2^{\prime} r_0(y_2);~~~y_1>y_2, 
	\end{split}
	\end{align}
	$\theta_{i},\theta_i^{\prime}>0;~i=1,2$. The parameters $\theta_{i},\theta_i^{\prime}>0;~i=1,2$ are the proportionality parameters which signify the change in the structural behaviour of distribution of lifetimes when the information on the failure time of the other component is given. These type of models suggest and are relevant when the failure of one component affects the stochastic structure of the lifetime of the surviving component.
	
	\begin{thm} 
		The underlying bivariate density function of $(Y_1,Y_2)$ satisfying (\ref{prhrd}) is 
	\begin{equation}\label{bdf}
		f_{Y_{1}, Y_{2}}\left(y_{1}, y_{2}\right)=\left\{\begin{array}{ll}{\theta_{1}^{\prime} \theta_{2} f_{0}\left(y_{1}\right) f_{0}\left(y_{2}\right)\left[F_{0}\left(y_{1}\right)\right]^{\theta_{1}^{\prime}-1}\left[F_{0}\left(y_{2}\right)\right]^{\theta_{1}+\theta_{2}-\theta_{1}^{\prime}-1}} & {; a<y_{1}<y_{2}<b} \\ {\theta_{1} \theta_{2}^{\prime} f_{0}\left(y_{1}\right) f_{0}\left(y_{2}\right)\left[F_{0}\left(y_{1}\right)\right]^{\theta_{1}+\theta_{2}-\theta_{2}^{\prime}-1}{\left[F_{0}\left(y_{2}\right)\right]^{\theta_{2}^{\prime}-1}}} & {; a<y_{2}<y_{1}<b}\end{array}\right.,
		\end{equation}
		for $\theta_{i},\theta_i^{\prime}>0;~\text{i=1,2}$. 
	\end{thm}
	
	\begin{proof}
		Substituting \eqref{prhrd} in \eqref{coxpdf}, the expression in \eqref{bdf} follows. It is straightforward to observe that
		
\begin{landscape}
	\begin{table}
		\begin{threeparttable}
			\caption{Members of the DPRH class \label{tabeg}}
			\begin{tabular}{c|l|l}
				\hline
				\textbf{No.} & \textbf{Baseline distribution} &  \textbf{Joint density function} \\ \hline
				\multirow{3}{*}{1.} & Exponentiated Gumbel & \multirow{3}{*}{$f_{Y_{1}, Y_{2}}\left(y_{1}, y_{2}\right)=\left\{\begin{array}{ll}{\theta_{1}^{\prime} \theta_{2} \lambda^{2} e^{-\lambda (y_1+y_2)} e^{-\theta_{1}^{\prime}e^{-\lambda y_1}} e^{-(\theta_1+\theta_2-\theta_{1}^{\prime})e^{-\lambda y_2}}} & {; y_{2}>y_{1}} \\
				{\theta_{1} \theta_{2}^{\prime} \lambda^{2} e^{-\lambda (y_1+y_2)} e^{-(\theta_1+\theta_2-\theta_{2}^{\prime})e^{-\lambda y_1}} e^{-\theta_{2}^{\prime}e^{-\lambda y_2}}} & {; y_{1}>y_{2}}\end{array}\right.$}  \\ & $F_{0}(y)=e^{-e^{-\lambda y}};$ & \\ 
				& $-\infty < y < \infty,~\lambda>0$ &  \\ \hline
				\multirow{3}{*}{2.} & Generalized Exponential & \multirow{3}{*}{$f_{Y_{1}, Y_{2}}\left(y_{1}, y_{2}\right)=\left\{\begin{array}{ll}{\theta_1^{\prime}\theta_2 \lambda^{2} e^{-\lambda(y_1+y_2)}(1-e^{-\lambda y_1})^{\theta_1^{\prime}-1}(1-e^{-\lambda y_2})^{\theta_1+\theta_2-\theta_1^{\prime}-1}} & {; y_{2}>y_{1}>0} \\
				{\theta_1 \theta_2^{\prime} \lambda^{2} e^{-\lambda (y_1+y_2)}(1-e^{-\lambda y_1})^{\theta_1+\theta_2-\theta_2^{\prime}-1}(1-e^{-\lambda y_2})^{\theta_2^{\prime}-1}} & {; y_{1}>y_{2}>0}\end{array}\right.$}  \\ & $F_0(y)=1-e^{-\lambda y};$ & \\ 
				& $y>0,~\lambda >0$ &  \\ \hline
				\multirow{4}{*}{3.} & Generalized Inverse Rayleigh & \multirow{3}{*}{$f_{Y_{1}, Y_{2}}\left(y_{1}, y_{2}\right)=\left\{\begin{array}{ll}{\frac{4\theta_{1}^{\prime} \theta_{2} \alpha^2 \lambda^2}{(y_1-\mu)^3 (y_2-\mu)^3} e^{-\lambda k}  (k_1 k_2 )^{\alpha-1} (1-k_1^\alpha)^{\theta_{1}^{\prime}-1} (1-k_2^\alpha)^{\theta_{1} +\theta_{2}-\theta_{1}^{\prime}-1}} & {; y_{2}>y_{1}>\mu} \\ {\frac{4\theta_{1} \theta_{2}^{\prime} \alpha^2 \lambda^2}{(y_1-\mu)^3 (y_2-\mu)^3} e^{-\lambda k}  (k_1 k_2 )^{\alpha-1} (1-k_1^\alpha)^{\theta_{1} +\theta_{2}-\theta_{2}^{\prime}-1} (1-k_2^\alpha)^{  \theta_{2}^{\prime}-1}} & {; y_{1}>y_{2}>\mu}\end{array}\right.$}  \\ & \multirow{2}{*}{$F_0(y)=1-(1-e^{-\frac{\lambda}{(y-\mu)^2}})^\alpha;$} & \\ 
				& & \\
				& $y>\mu,~\alpha>0,~\lambda>0$  & where $k=\frac{1}{(y_1-\mu)^2}+\frac{1}{(y_2-\mu)^2}$ and $k_i=1-e^{-\frac{\lambda}{(y_i-\mu)^2}};~i=1,2$ \\ \hline
				\multirow{5}{*}{4.} & Generalized Rayleigh & \multirow{5}{*}{$f_{Y_{1}, Y_{2}}\left(y_{1}, y_{2}\right)=\left\{\begin{array}{ll}{4\theta_1^{\prime}\theta_2 \alpha^{2} \lambda^{4} y_1 y_2 e^{-\lambda^{2}(y_1^2+y_2^2)}(1-e^{-(\lambda y_1)^2})^{\alpha \theta_1^{\prime}-1}(1-e^{-(\lambda y_2)^2})^{\alpha (\theta_1+\theta_2-\theta_1^{\prime})-1}}\\ \hspace{9cm} {; y_{2}>y_{1}>0} \\
				{4\theta_1 \theta_2^{\prime} \alpha^{2} \lambda^{4} y_1 y_2 e^{-\lambda^{2}(y_1^2+y_2^2)}(1-e^{-(\lambda y_1)^2})^{\alpha (\theta_1+\theta_2-\theta_2^{\prime})-1}(1-e^{-(\lambda y_2)^2})^{\alpha \theta_2^{\prime}-1}} \\ \hspace{9cm}{; y_{1}>y_{2}>0}\end{array}\right.$}  \\ & \multirow{3}{*}{$F_0(y)=(1-e^{-(\lambda y)^2})^{\alpha};$} & \\ 
				&  &  \\
				&  &  \\
				& $y>0,~\lambda>0,~\alpha>0$ &  \\ \hline
				\multirow{4}{*}{5.} & Inverse Exponential & \multirow{4}{*}{$f_{Y_{1}, Y_{2}}\left(y_{1}, y_{2}\right)=\left\{\begin{array}{ll}{\frac{\theta_{1}^{\prime}\theta_2 \lambda^{2}}{y_{1}^{2} y_{2}^{2}}e^{-\frac{\lambda \theta_1^{\prime}}{y_1}}e^{-\frac{\lambda}{y_2}(\theta_1+\theta_2-\theta_1^{\prime})}~~~\text{ if }~y_{2}>y_{1}>0 }\\
				{\frac{\theta_{1}\theta_2^{\prime}\lambda^{2}}{y_{1}^{2} y_{2}^{2}}e^{-\frac{\lambda}{y_1}(\theta_1+\theta_2-\theta_2^{\prime})}e^{-\frac{\lambda \theta_2^{\prime}}{y_2}}~~~\text{ if }~y_{1}>y_{2}>0 }\end{array}\right.$}  \\ & \multirow{2}{*}{$F_0(y)=e^{-\frac{\lambda}{y}};$} & \\ 
				&  &  \\
				& $y>0,~\lambda>0$ &  \\ \hline
				\multirow{4}{*}{6.} & Burr Type III & \multirow{4}{*}{$f_{Y_{1}, Y_{2}}\left(y_{1}, y_{2}\right)=\left\{\begin{array}{ll}{\frac{c^2 \theta_1^{\prime}\theta_2 (y_1 y_2)^{-(c+1)}}{(1+y_1^{-c})^{\theta_1^{\prime}+1} (1+y_2^{-c})^{\theta_1+\theta_2-\theta_1^{\prime}+1}}} & {\text { if } y_{2}>y_{1}>0} \\
				{\frac{c^2 \theta_1 \theta_2^{\prime} (y_1 y_2)^{-(c+1)}}{ (1+y_1^{-c})^{\theta_1+\theta_2-\theta_2^{\prime}+1}(1+y_2^{-c})^{\theta_2^{\prime}+1}}} & {\text { if } y_{1}>y_{2}>0}\end{array}\right.$}  \\ & \multirow{2}{*}{$F_0(y)=\frac{1}{1+y^{-c}};$} & \\ 
				&  &  \\
				& $y>0,~c>0$ &  \\ \hline
				\multirow{3}{*}{7.} & Inverse Weibull & \multirow{3}{*}{$f_{Y_{1}, Y_{2}}\left(y_{1}, y_{2}\right)=\left\{\begin{array}{ll}{\theta_1^{\prime}\theta_2\alpha^2(y_1 y_2)^{-(\alpha+1)} (e^{-y_1^{-\alpha}})^{\theta_1^{\prime}} (e^{-y_2^{-\alpha}})^{\theta_1+\theta_2-\theta_1^{\prime}}} & {\text { if } y_{2}>y_{1}>0} \\ {\theta_1\theta_2^{\prime}\alpha^2(y_1 y_2)^{-(\alpha+1)}(e^{-y_1^{-\alpha}})^{\theta_1+\theta_2-\theta_2^{\prime}}(e^{-y_2^{-\alpha}})^{\theta_2^{\prime}}} & {\text { if } y_{1}>y_{2}>0}\end{array}\right.$}  \\ & $F_0(y)=e^{-y^{-\alpha}};$ & \\ 
				& $y>0,~\alpha>0.$ &  \\ \hline
			\end{tabular}
			\begin{tablenotes}
				\small
				\item where $\theta_i,~\theta_{i}^{\prime}>0;~i=1,2$.
			\end{tablenotes}
		\end{threeparttable}
	\end{table}
\end{landscape}
		
		$f(y_1,y_2) \geq 0$ for all $a<y_1,y_2<b$ and
		\begin{align*}
		\int_{y_1} \int_{y_2} f(y_1,y_2)dy_2 dy_1=&\int_{a}^{b} \int_{a}^{y_1} \theta_{1} \theta_{2}^{\prime} f_{0}\left(y_{1}\right) f_{0}\left(y_{2}\right)\left[F_{0}\left(y_{1}\right)\right]^{\theta_{1}+\theta_{2}-\theta_{2}^{\prime}-1}{\left[F_{0}\left(y_{2}\right)\right]^{\theta_{2}^{\prime}-1}} dy_2 dy_1 \\
		&+ \int_{a}^{b} \int_{y_1}^{b} \theta_{1}^{\prime} \theta_{2} f_{0}\left(y_{1}\right) f_{0}\left(y_{2}\right)\left[F_{0}\left(y_{1}\right)\right]^{\theta_{1}^{\prime}-1}\left[F_{0}\left(y_{2}\right)\right]^{\theta_{1}+\theta_{2}-\theta_{1}^{\prime}-1} dy_2 dy_1 \\
		=&\frac{\theta_1}{\theta_1+\theta_2} + \frac{\theta_2}{\theta_1+\theta_2}=1.
		\end{align*}
	\end{proof}
	 Since the density form differs for $y_1 < y_2$ and $y_1 > y_2$ we refer to the bivariate class of distributions in \eqref{bdf} as \textit{dynamic proportional reversed hazards (\textit{DPRH}) model} and denote it by $DPRH(F_{0},\theta_{1},\theta_{2},\theta_{1}^{\prime},\theta_{2}^{\prime})$ where $F_0$ is the baseline distribution. 
	\begin{cor} 
		A necessary and sufficient condition under which $Y_1$ and $Y_2$ are independent is $\theta_{i}^{\prime}=\theta_i,\, \text{i=1,2}.$
	\end{cor}
	\par
\begin{figure}	
		\centering
		\begin{minipage}{0.45\textwidth}\label{fig0012}
			\includegraphics[width=0.9\textwidth]{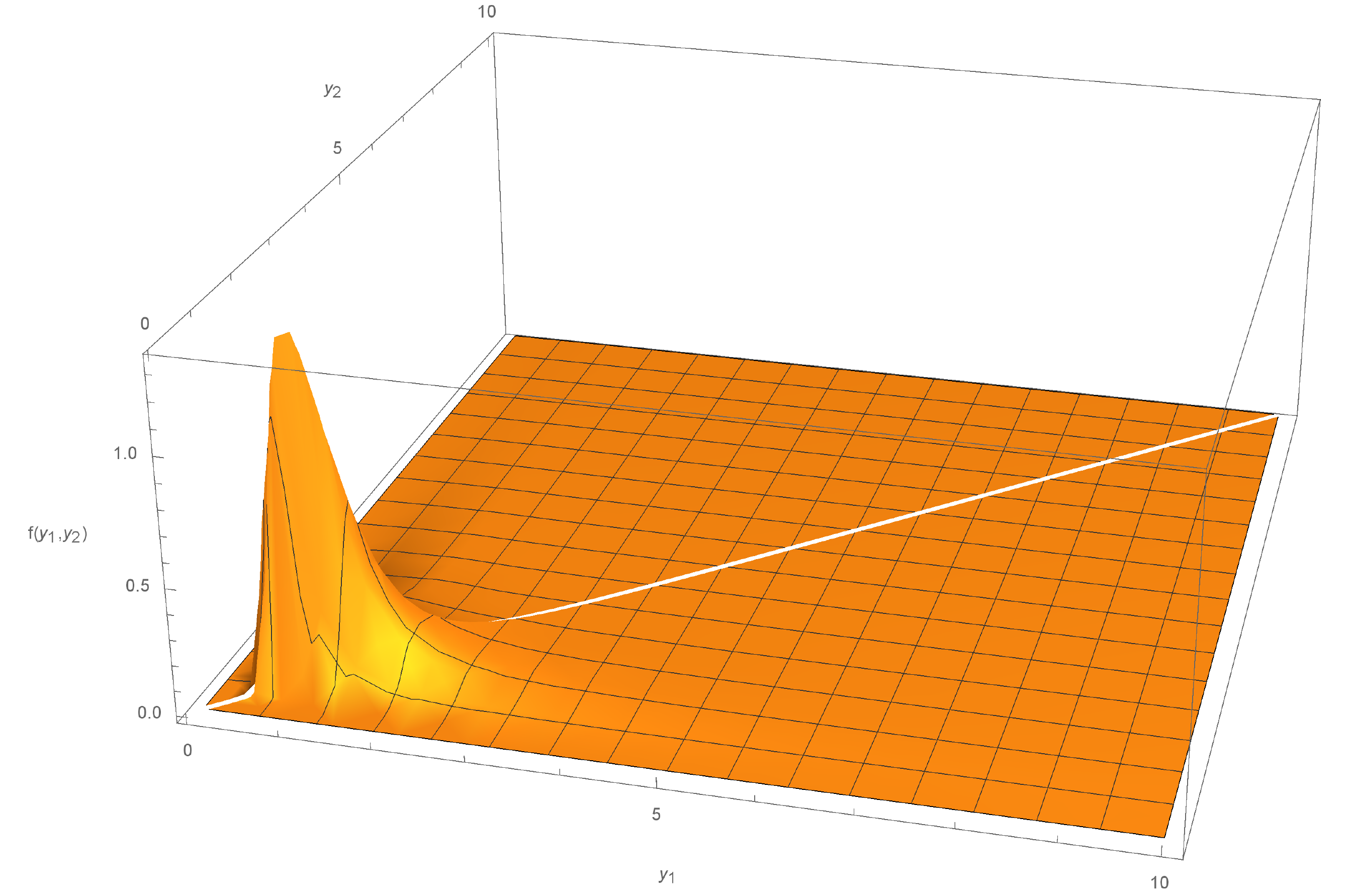} 
			\subcaption{DPRH(IW(2),1,1,2,3)}
		\end{minipage}\hfill
		\begin{minipage}{0.45\textwidth}\label{fig0016}
			\centering
			\includegraphics[width=0.9\textwidth]{Figures/Fig1} 
			\subcaption{DPRH(IW(3),1,1,2.1,2.5)}
		\end{minipage}	
		\caption{Plots of the joint pdf the \textit{DPRH} model with Inverse Weibull as the baseline distribution}
		\label{figiw}
	\end{figure}

 Some members of the \textit{DPRH} class are given in Table \ref{tabeg}. Figure \ref{figiw} shows the plot of the joint pdf with baseline distribution as Inverse Weibull for different parameter values.
	
\section{Properties} \label{properties}
The cumulative distribution function for various possibilities of the parameters are derived in the next theorem.
\begin{thm} \label{pptydf}
	Let $(Y_1,Y_2) \sim DPRH(F_0,\theta_{1},\theta_{2},\theta_{1}^{\prime},\theta_{2}^{\prime})$. Then the cumulative distribution function of $(Y_1,Y_2)$, $F(y_1,y_2)$, in each case is\\\\
	\textit{Case 1:} $\theta_1+\theta_2 \neq \theta_1^{\prime}$ and $\theta_1+\theta_2 \neq \theta_2^{\prime}$.
	\begin{equation} \label{df1}
		F\left(y_{1}, y_{2}\right)=\left\{\begin{array}{ll}{[F_0(y_1)]^{\theta_1+\theta_2}+\frac{\theta_2 [F_0(y_1)]^{\theta_1^{\prime}}}{\theta_1+\theta_2-\theta_1^{\prime}} \Big[[F_0(y_2)]^{\theta_1+\theta_2-\theta_1^{\prime}}-[F_0(y_1)]^{\theta_1+\theta_2-\theta_1^{\prime}} \Big]~~~\text{ if }~y_{1}<y_{2} }\\{[F_0(y_2)]^{\theta_1+\theta_2}+\frac{\theta_1 [F_0(y_2)]^{\theta_2^{\prime}}}{\theta_1+\theta_2-\theta_2^{\prime}} \Big[[F_0(y_1)]^{\theta_1+\theta_2-\theta_2^{\prime}}-[F_0(y_2)]^{\theta_1+\theta_2-\theta_2^{\prime}} \Big]~~~\text{ if }~y_{1}>y_{2} }\end{array}\right.
	\end{equation}	
	\textit{Case 2:} $\theta_1+\theta_2 = \theta_1^{\prime}$ and $\theta_1+\theta_2 \neq \theta_2^{\prime}$.
	\begin{equation} \label{df2}
		F\left(y_{1}, y_{2}\right)=\left\{\begin{array}{ll}{[F_0(y_1)]^{\theta_1+\theta_2} \Big [1+ \theta_2 ln \Big ( \frac{F_0(y_2)}{F_0(y_1)}\Big ) \Big ]~~~\text{ if }~y_{1}<y_{2} }\\{[F_0(y_2)]^{\theta_1+\theta_2}+\frac{\theta_1 [F_0(y_2)]^{\theta_2^{\prime}}}{\theta_1+\theta_2-\theta_2^{\prime}} \Big[[F_0(y_1)]^{\theta_1+\theta_2-\theta_2^{\prime}}-[F_0(y_2)]^{\theta_1+\theta_2-\theta_2^{\prime}} \Big]~~~\text{ if }~y_{1}>y_{2} }\end{array}\right.
	\end{equation}
	\textit{Case 3:} $\theta_1+\theta_2 \neq \theta_1^{\prime}$ and $\theta_1+\theta_2 = \theta_2^{\prime}$.
	\begin{equation} \label{df3}
		F\left(y_{1}, y_{2}\right)=\left\{\begin{array}{ll}{[F_0(y_1)]^{\theta_1+\theta_2}+\frac{\theta_2 [F_0(y_1)]^{\theta_1^{\prime}}}{\theta_1+\theta_2-\theta_1^{\prime}} \Big[[F_0(y_2)]^{\theta_1+\theta_2-\theta_1^{\prime}}-[F_0(y_1)]^{\theta_1+\theta_2-\theta_1^{\prime}} \Big]~~~\text{ if }~y_{1}<y_{2} }\\{[F_0(y_2)]^{\theta_1+\theta_2} \Big [1+ \theta_1 ln \Big ( \frac{F_0(y_1)}{F_0(y_2)}\Big ) \Big ]~~~\text{ if }~y_{1}>y_{2} }\end{array}\right.
	\end{equation}
	\textit{Case 4:} $\theta_1+\theta_2 = \theta_1^{\prime} = \theta_2^{\prime}$.
	\begin{equation} \label{df4}
		F\left(y_{1}, y_{2}\right)=\left\{\begin{array}{ll}{[F_0(y_1)]^{\theta_1+\theta_2} \Big [1+ \theta_2 ln \Big ( \frac{F_0(y_2)}{F_0(y_1)}\Big ) \Big ]~~~\text{ if }~y_{1}<y_{2} }\\{[F_0(y_2)]^{\theta_1+\theta_2} \Big [1+ \theta_1 ln \Big ( \frac{F_0(y_1)}{F_0(y_2)}\Big ) \Big ]~~~\text{ if }~y_{1}>y_{2} }\end{array}\right.
	\end{equation}
\end{thm}
	
	\begin{thm}
		Let $(Y_1,Y_2) \sim DPRH(F_0,\theta_{1},\theta_{2},\theta_{1}^{\prime},\theta_{2}^{\prime})$. Then $Y=\text{max}\{Y_1,Y_2\}$ is a PRH model.
	\end{thm}
	
	\begin{proof}
		We have,
		\begin{align*}
		P(Y \leq y)=&F(y,y) \\
		=&[F_0(y)]^{\theta_1+\theta_2}+\frac{\theta_2 [F_0(y)]^{\theta_1^{\prime}}}{\theta_1+\theta_2-\theta_1^{\prime}} \Big [ [F_0(y)]^{\theta_1+\theta_2-\theta_1^{\prime}}- [F_0(y)]^{\theta_1+\theta_2-\theta_1^{\prime}} \Big] \\
		=&[F_0(y)]^{\theta_1+\theta_2}.
		\end{align*}
	\end{proof}
	
	\begin{thm}
		Let $(Y_1,Y_2) \sim DPRH(F_0,\theta_{1},\theta_{2},\theta_{1}^{\prime},\theta_{2}^{\prime})$. Then $P(Y_i>Y_{3-i})=\frac{\theta_i}{\theta_i+\theta_{3-i}},~\text{i=1,2}$.
	\end{thm}
	
	\begin{proof}
		From \eqref{bdf}, we have, for $i=1,2$,
		\begin{align*}
		P(Y_i>Y_{3-i})=&\int_{0}^{\infty} \int_{0}^{y_i} f_{(y_i>y_{3-i})}(y_1,y_2) dy_{3-i} dy_i \\
		=&\theta_i \theta_{3-i}^{\prime} \int_{0}^{\infty} f_0(y_i)[F_0(y_i)]^{\theta_i+\theta_{3-i}-\theta_{3-i}^{\prime}-1} \int_{0}^{y_i} f_0(y_{3-i})[F_0(y_{3-i})]^{\theta_{3-i}^{\prime}-1} dy_{3-i} dy_i \\
		=& \theta_i \int_{0}^{\infty} f_0(y_i)[F_0(y_i)]^{\theta_i+\theta_{3-i}-1} dy_i \\
		=&\frac{\theta_i}{\theta_i+\theta_{3-i}}.
		\end{align*}
		Hence the proof.
	\end{proof}
	
	\begin{thm}
		Let $(Y_1,Y_2) \sim DPRH(F_0,\theta_{1},\theta_{2},\theta_{1}^{\prime},\theta_{2}^{\prime})$. Then the marginal distribution functions of $Y_1$ and $Y_2$ are\\\\
		\textit{Case 1:} $\theta_1+\theta_2 \neq \theta_1^{\prime}$ and $\theta_1+\theta_2 \neq \theta_2^{\prime}$.
		\begin{equation*}
			F_{Y_i}(y_i)=\frac{\theta_{3-i}}{\theta_i+\theta_{3-i}-\theta_i^{\prime}} [F_0(y_i)]^{\theta_i^{\prime}}+ \frac{\theta_i-\theta_i^{\prime}}{\theta_i+\theta_{3-i}-\theta_i^{\prime}}[F_0(y_i)]^{\theta_i+\theta_{3-i}}~~~\text{if}~~a<y_{i}<b;~i=1,2.
		\end{equation*}
		\textit{Case 2:} $\theta_1+\theta_2 = \theta_1^{\prime}$ and $\theta_1+\theta_2 \neq \theta_2^{\prime}$.
		\begin{align*}
			\begin{split}
				F_{Y_1}(y_1)=&[F_0(y_1)]^{\theta_1+\theta_2}[1-\theta_2 ln(F_0(y_1))]~~~\text{if}~~a<y_{1}<b \\
				F_{Y_2}(y_2)=&\frac{\theta_1}{\theta_1+\theta_2-\theta_2^{\prime}} [F_0(y_2)]^{\theta_2^{\prime}}+ \frac{\theta_2-\theta_2^{\prime}}{\theta_1+\theta_2-\theta_2^{\prime}}[F_0(y_2)]^{\theta_1+\theta_2}~~~\text{if}~~a<y_{2}<b.
			\end{split}
		\end{align*}
		\textit{Case 3:} $\theta_1+\theta_2 \neq \theta_1^{\prime}$ and $\theta_1+\theta_2 = \theta_2^{\prime}$.
		\begin{align*}
			\begin{split}
				F_{Y_1}(y_1)=&\frac{\theta_2}{\theta_1+\theta_2-\theta_1^{\prime}} [F_0(y_1)]^{\theta_1^{\prime}}+ \frac{\theta_1-\theta_1^{\prime}}{\theta_1+\theta_2-\theta_1^{\prime}}[F_0(y_1)]^{\theta_1+\theta_2}~~~\text{if}~~a<y_{1}<b \\
				F_{Y_2}(y_2)=&[F_0(y_2)]^{\theta_1+\theta_2}[1-\theta_1 ln(F_0(y_2))]~~~\text{if}~~a<y_{2}<b.
			\end{split}
		\end{align*}
		\textit{Case 4:} $\theta_1+\theta_2 = \theta_1^{\prime}$ and $\theta_1+\theta_2 = \theta_2^{\prime}$.
		\begin{equation*}
			F_{Y_i}(y_i)=[F_0(y_i)]^{\theta_i+\theta_{3-i}}[1-\theta_{3-i} ln(F_0(y_i))]~~~\text{if}~~a<y_{i}<b;~i=1,2.
		\end{equation*}
	\end{thm}
	
	\begin{proof}
		\textit{Case 1:} $\theta_1+\theta_2 \neq \theta_1^{\prime}$ and $\theta_1+\theta_2 \neq \theta_2^{\prime}$. \vspace{0.5cm}\\
		For $a<y_1<b$, from \eqref{df1}
		\begin{align*}
		F_{Y_1}(y_1)=&F(y_1,\infty)\\
		=&[F_0(y_1)]^{\theta_1+\theta_2}+\frac{\theta_2 [F_0(y_1)]^{\theta_1^{\prime}}}{\theta_1+\theta_2-\theta_1^{\prime}} \Big[1-[F_0(y_1)]^{\theta_1+\theta_2-\theta_1^{\prime}} \Big]\\
		=&\frac{\theta_2}{\theta_1+\theta_2-\theta_1^{\prime}} [F_0(y_1)]^{\theta_1^{\prime}}+ \frac{\theta_1-\theta_1^{\prime}}{\theta_1+\theta_2-\theta_1^{\prime}}[F_0(y_1)]^{\theta_1+\theta_2}.
		\end{align*}
		Similarly, for $a<y_2<b$, we get,
		\begin{equation*}
		F_{Y_2}(y_2)=\frac{\theta_1}{\theta_1+\theta_2-\theta_2^{\prime}} [F_0(y_2)]^{\theta_2^{\prime}}+ \frac{\theta_2-\theta_2^{\prime}}{\theta_1+\theta_2-\theta_2^{\prime}}[F_0(y_2)]^{\theta_1+\theta_2}.
		\end{equation*}
		\textit{Case 2:} $\theta_1+\theta_2 = \theta_1^{\prime}$ and $\theta_1+\theta_2 \neq \theta_2^{\prime}$. \vspace{0.5cm}\\
		For $a<y_2<b$, $F_{Y_2}(y_2)$ follows as in \textit{Case 1}. For $a<y_1<b$, from \eqref{df2} it follows that 
		\begin{equation*}
		F_{Y_1}(y_1)=[F_0(y_1)]^{\theta_1+\theta_2}[1-\theta_2 ln(F_0(y_1))].
		\end{equation*}
		\textit{Case 3:} $\theta_1+\theta_2 \neq \theta_1^{\prime}$ and $\theta_1+\theta_2 = \theta_2^{\prime}$. \vspace{0.5cm}\\
		For $a<y_1<b$, $F_{Y_1}(y_1)$ follows as in \textit{Case 1}. For $a<y_2<b$, from \eqref{df3} it follows that 
		\begin{equation*}
		F_{Y_2}(y_2)=[F_0(y_2)]^{\theta_1+\theta_2}[1-\theta_1 ln(F_0(y_2))].
		\end{equation*}
		\textit{Case 4:} $\theta_1+\theta_2 = \theta_1^{\prime}$ and $\theta_1+\theta_2 = \theta_2^{\prime}$.\vspace{0.5cm} \\
		For $a<y_1<b$, $F_{Y_1}(y_1)$ follows as in \textit{Case 2} and for $a<y_2<b$, $F_{Y_2}(y_2)$ follows as in \textit{Case 3}.\\
		Hence the proof.
	\end{proof}
	
	\begin{thm}
		Let $(Y_1,Y_2) \sim DPRH(F_0,\theta_{1},\theta_{2},\theta_{1}^{\prime},\theta_{2}^{\prime})$. If $\theta_1^{\prime}+\theta_2^{\prime}>\theta_1+\theta_2$, then $(Y_1,Y_2)$ has total positivity of order two $(TP2)$ property. 
	\end{thm}
	
	\begin{proof}
		$(Y_1,Y_2)$ has total positivity of order two $(TP2)$ property iff for any $y_{11},y_{12},y_{21},y_{22}$, whenever $0<y_{11}<y_{12}$ and $0<y_{21}<y_{22}$, we have
		\begin{equation} \label{tp2}
		f(y_{11},y_{21})f(y_{12},y_{22}) \geq f(y_{12},y_{21})f(y_{11},y_{22}).
		\end{equation}
		Consider the case, $y_{11}<y_{21}<y_{12}<y_{22}$. Then proving \eqref{tp2} is equivalent to proving 
		\begin{equation*}
		[F_0(y_{12})]^{\theta_1^{\prime}+\theta_2^{\prime}-\theta_1-\theta_2} \geq [F_0(y_{21})]^{\theta_1^{\prime}+\theta_2^{\prime}-\theta_1-\theta_2}
		\end{equation*}
		which is always true for $\theta_1^{\prime}+\theta_2^{\prime}>\theta_1+\theta_2$ and $y_{12}>y_{21}$. Similarly, for all the remaining possible combinations of $y_{11},y_{12},y_{21}$ and $y_{22}$, \eqref{tp2} can be proved. Hence, $(Y_1,Y_2)$ has $TP2$ property if $\theta_1^{\prime}+\theta_2^{\prime}>\theta_1+\theta_2$.
	\end{proof}
	
	\citet{pg2006bivariate} introduced a new local dependence measure which describes the time-varying dependence between $Y_1$ and $Y_2$ in terms of reversed hazard rate and is defined at $(y_1,y_2)$ by,
	\begin{equation*}
	\beta(y_1,y_2)=\frac{F\left(y_{1}, y_{2}\right) f\left(y_{1}, y_{2}\right)}{F_{1}\left(y_{1}, y_{2}\right) F_{2}\left(y_{1}, y_{2}\right)},
	\end{equation*}
	where $F_{1}\left(y_{1}, y_{2}\right)=\frac{\partial F(y_1,y_2)}{\partial y_1}$ and $F_{2}\left(y_{1}, y_{2}\right)=\frac{\partial F(y_1,y_2)}{\partial y_2}$. The measure $\beta(y_1,y_2)$ is the ratio of the reversed hazard rates of the conditional distribution of $Y_{i}$ given $Y_{3-i}=y_{3-i}$ to that of $Y_{i}$ given $Y_{3-i}<y_{3-i}$ for $i=1,2$. It is equal to $1$ if and only if $Y_{1}$ and $Y_{2}$ are independent.\\
	The local dependence measure for \textit{DPRH} model is given by,
	\begin{equation*} \label{crf1}
	\beta(y_1,y_2)=\frac{\theta_i^{\prime}(\theta_i - \theta_i^{\prime})[F_0(y_i)]^{\theta_i+\theta_{3-i}-\theta_i^{\prime}}+\theta_i^{\prime}\theta_{3-i} [F_0(y_{3-i})]^{\theta_i+\theta_{3-i}-\theta_i^{\prime}}}{(\theta_i - \theta_i^{\prime})(\theta_i+\theta_{3-i})[F_0(y_i)]^{\theta_i+\theta_{3-i}-\theta_i^{\prime}}+\theta_i^{\prime}\theta_{3-i} [F_0(y_{3-i})]^{\theta_i+\theta_{3-i}-\theta_i^{\prime}}},
	\end{equation*}
	for $y_i<y_{3-i}$ and $\theta_1+\theta_2 \neq \theta_i^{\prime};~i=1,2$ and 
	\begin{equation*} \label{crf2}
		\beta(y_1,y_2)=\frac{\theta_i^{\prime}+\theta_{3-i} \theta_{i}^{\prime}log \big (\frac{F_{0}(y_{3-i})}{F_{0}(y_i)}\big )}{\theta_i+\theta_{3-i} \theta_{i}^{\prime}log \big (\frac{F_{0}(y_{3-i})}{F_{0}(y_i)}\big )},
	\end{equation*}
	for $y_i<y_{3-i}$ and $\theta_1+\theta_2 = \theta_i^{\prime};~i=1,2$.
	\begin{thm}
		The local dependence measure $\beta(y_1,y_2)=1$ if and only if $\theta_1=\theta_{1}^{\prime}$ and $\theta_2=\theta_{2}^{\prime}$.
	\end{thm}
	
	\par The issue of identifiability of parameters often pose a grave problem in estimation of parameters of a model. The next theorem investigates the identifiability of the \textit{DPRH} model. We first state the definition of identifiability.
	
	\begin{dfntn}
		If $f_{\Theta_1}$ and $f_{\Theta_2}$ are two members of $\mathscr{F}$, a class of distributions, with $\Theta_i$ as corresponding vector of parameters, then $\mathscr{F}$ is said to be a class of identifiable distributions if  $f_{\Theta_1}=f_{\Theta_2}$ implies $\Theta_1=\Theta_2$, for every vector $\Theta_i$ where equality implies component wise equality.
	\end{dfntn}
	
	\begin{thm}
		The \textit{DPRH} model is identifiable for an identifiable baseline distribution.
	\end{thm}
	
	\begin{proof}
		For every $y_1<y_2$,  $f_{\Theta_1} = f_{\Theta_2}$
		\begin{equation} \label{iden} 
			\implies \theta_{1}^{\prime} \theta_{2} \left[F_{0}\left(y_{1}\right)\right]^{\theta_{1}^{\prime}-\beta_{1}^{\prime}}=\beta_{1}^{\prime} \beta_{2} \left[F_{0}\left(y_{2}\right)\right]^{(\beta-\beta_{1}^{\prime})-(\theta-\theta_{1}^{\prime})},~\text{for all}~y_1, y_2,
		\end{equation}
		where $\theta=\theta_1+\theta_2$ and $\beta=\beta_1+\beta_2$. Since LHS is a function of $y_1 $ only and RHS is a function $y_2$ only, it in turn implies that  $\left[F_{0}\left(y_{1}\right)\right]^{\theta_{1}^{\prime}-\beta_{1}^{\prime}}$ and $\left[F_{0}\left(y_{2}\right)\right]^{(\beta-\beta_{1}^{\prime})-(\theta-\theta_{1}^{\prime})}$ are identically equal to unity. Hence, $\theta_1^{\prime}=\beta_1^{\prime}$ and $\theta_2=\beta_2$. Similarly arguing for $y_1>y_2$, we get $\theta_2^{\prime}=\beta_2^{\prime}$ and $\theta_1=\beta_1$.
	\end{proof}
	
	\section{Estimation of Parameters} \label{estimation}
	In this section, we discuss the maximum likelihood estimation and the Bayesian inference of the unknown parameters in the \textit{DPRH} model based on a random sample of size $n$.
	\subsection{Maximum Likelihood Estimation}\label{mle}
	Let the random sample be $\{(y_{1k},y_{2k}); k=1,2,\ldots,n\}$. For a complete data, define the sets $I_1$ and $I_2$ as $I_1=\{k;~y_{1k}>y_{2k}\}$ and $I_2=\{k;~y_{1k}<y_{2k}\}$ such that $I=I_1 \cup I_2$. Let $m_1$ and $m_2$ be the cardinality of $I_1$ and $I_2$, respectively, such that $m_1+m_2=n,~m_1,m_2>0$ and $\Theta=(\theta_1,\theta_2,\theta_1^{\prime},\theta_2^{\prime})$. The log-likelihood function based on the observations is,
	\begin{align}
		\nonumber \log L(\Theta)=&m_1 \ln \theta_1 +\theta_1 \left(\sum_{k \in I_{1}} \ln \left(F_{0}\left(y_{1 k}\right)\right)+\sum_{k \in I_{2}} \ln \left(F_{0}\left(y_{2 k}\right)\right)\right)\\ \nonumber
		&+m_2 \ln \theta_2 +\theta_2 \left(\sum_{k \in I_{1}} \ln \left(F_{0}\left(y_{1 k}\right)\right)+\sum_{k \in I_{2}} \ln \left(F_{0}\left(y_{2 k}\right)\right)\right)\\ \nonumber
		&+m_{2} \ln \theta_{1}^{\prime}+\theta_{1}^{\prime}\left(\sum_{k \in I_{2}} \ln \left(F_{0}\left(y_{1 k}\right)\right)-\sum_{k \in I_{2}} \ln \left(F_{0}\left(y_{2 k}\right)\right)\right)\\ \nonumber
		&+m_{1} \ln \theta_{2}^{\prime}+\theta_{2}^{\prime}\left(\sum_{k \in I_{1}} \ln \left(F_{0}\left(y_{2 k}\right)\right)-\sum_{k \in I_{1}} \ln \left(F_{0}\left(y_{1 k}\right)\right)\right)\\ \nonumber
		&+\sum_{k \in I} \ln f_{0}\left(y_{1 k}\right)+\sum_{k \in I} \ln f_{0}\left(y_{2 k}\right)-\sum_{k \in I} \ln \left(F_{0}\left(y_{1 k}\right)\right)-\sum_{k \in I} \ln \left(F_{0}\left(y_{2 k}\right)\right).
	\end{align}
	The maximum likelihood estimates (MLE) of $\theta_1,\theta_2,\theta_1^{\prime},\theta_2^{\prime}$ are in a closed form as follows.
	\begin{eqnarray} \nonumber
		\begin{split}
			&\hat{\theta_i}=-\frac{m_i}{\sum_{k \in I_{1}} \ln F_{0}\left(y_{1 k}\right)+\Sigma_{k \in I_{2}} \ln F_{0}\left(y_{2 k}\right)};~i=1,2\\
			&\hat{\theta}_{i}^{\prime}=\frac{m_{3-i}}{\sum_{k \in I_{3-i}} \ln F_{0}\left(y_{3-i,k}\right)-\Sigma_{k \in I_{3-i}} \ln F_{0}\left(y_{i k}\right)};~i=1,2.\\
		\end{split}
	\end{eqnarray}
	However if the baseline distributions are of the form $F_0(y; \pmb{\eta})$ where $\pmb{\eta}=(\eta_1,\eta_2,\ldots,\eta_p)$ are the vector of parameters in the baseline distribution, the MLE are not as easy to obtain. For all ensuing discussions, we assume $\pmb{\eta}=\eta$. Let $\Lambda=(\theta_1,\theta_2,\theta_1^{\prime},\theta_2^{\prime},\eta)$ be the vector of unknown parameters. Under left censoring, let $c_{1i}$ and $c_{2i}$ be the corresponding observed censoring times. Assume that the lifetimes and the censoring times are independent. The lifetime associated with the $r^{th}$ pair of components is,
	\begin{equation*}\label{lifetimecases}
	(Y_{1k},Y_{2k})=\left\{\begin{array}{ll}(y_{1k},y_{2k})~~~\text{if both $y_{1k}$ and $y_{2k}$ are uncensored}\\
	(y_{1k},c_{2k})~~~\text{if $y_{1k}$ is uncensored and $y_{2k}$ is censored}\\
	(c_{1k},y_{2k})~~~\text{if $y_{1k}$ is censored and $y_{2k}$ is uncensored}\\
	(c_{1k},c_{2k})~~~\text{if both $y_{1k}$ and $y_{2k}$ are censored} \end{array}\right.
	\end{equation*}
	Consider the following sets.
	\begin{align*}
		I_{j}&=\{k;~y_{jk} > c_{jk},~y_{3-j,k} > c_{3-j,k},~y_{jk} > y_{3-j,k}\};~j=1,2 \\
		I_{j+2}&=\{k;~y_{jk} > c_{jk},~y_{3-j,k} < c_{3-j,k},~y_{jk} > c_{3-j,k}\};~j=1,2 \\
		I_{j+4}&=\{k;~y_{jk} > c_{jk},~y_{3-j,k} < c_{3-j,k},~y_{jk} < c_{3-j,k}\};~j=1,2 \\
		I_{j+6}&=\{k;~y_{jk} < c_{jk},~y_{3-j,k} < c_{3-j,k},~c_{jk} > c_{3-j,k}\};~j=1,2.
	\end{align*}
	Let $n_{j+j^{\prime}}$ be the cardinality of $I_{j+j^{\prime}}$ for $j=1,2;~j^{\prime}=0,2,4,6$. The log-likelihood function given the data is obtained as in \eqref{likelihood} given in Appendix \ref{app1}. The likelihood equation is non-linear in nature and the maximum likelihood estimates are obtained by optimisation techniques in $R$. Since this is a regular family of distributions, the MLE will be asymptotically normally distributed as $N(\Lambda, \Sigma)$ (\cite{lehmann2006theory}), where $\Sigma$ is estimated as
	\begin{equation}\label{covmatrix}
		\hat{\Sigma}=\begin{bmatrix}
			\text{var}(\hat{\theta_{1}}) & \text{cov}(\hat{\theta_{1}},\hat{\theta_{2}}) & \text{cov}(\hat{\theta_{1}},\hat{\theta_{1}^{\prime}}) & \text{cov}(\hat{\theta_{1}},\hat{\theta_{2}^{\prime}}) & \text{cov}(\hat{\theta_{1}},\hat{\eta})\\  & \text{var}(\hat{\theta_{2}}) & \text{cov}(\hat{\theta_{2}},\hat{\theta_{1}^{\prime}}) & \text{cov}(\hat{\theta_{2}},\hat{\theta_{2}^{\prime}}) & \text{cov}(\hat{\theta_{2}},\hat{\eta}) \\ & & \text{var}(\hat{\theta_{1}^{\prime}}) & \text{cov}(\hat{\theta_{1}^{\prime}},\hat{\theta_{2}^{\prime}}) & \text{cov}(\hat{\theta_{1}^{\prime}},\hat{\eta}) \\ & & & \text{var}(\hat{\theta_{2}^{\prime}}) & \text{cov}(\hat{\theta_{2}^{\prime}},\hat{\eta}) \\ & & & & \text{var}(\hat{\eta})
		\end{bmatrix}=\begin{bmatrix}
			V_{11} & V_{12} & V_{13} & V_{14} & V_{15}\\  & V_{22} & V_{23} & V_{24} & V_{25} \\ & & V_{33} & V_{34} & V_{35} \\ & & &V_{44} & V_{45} \\ & & & & V_{55}
		\end{bmatrix}^{-1},
	\end{equation}
	with
	\begin{equation*}
		V_{ii}=\begin{cases}
			-\frac{\partial ^2 \log(L(\Lambda))}{\partial \theta_{i}^{2}} &;~i=1,2\\
			-\frac{\partial ^2 \log(L(\Lambda))}{\partial (\theta_{1}^{\prime})^{2}} &;~i=3\\
			-\frac{\partial ^2 \log(L(\Lambda))}{\partial (\theta_{2}^{\prime})^{2}} &;~i=4,
			\end{cases},~V_{i5}=\begin{cases}
			-\frac{\partial ^2 \log(L(\Lambda))}{\partial \theta_{i} \partial  \eta} &;~i=1,2\\
			-\frac{\partial ^2 \log(L(\Lambda))}{\partial \theta_{1}^{\prime} \partial  \eta} &;~i=3\\
			-\frac{\partial ^2 \log(L(\Lambda))}{\partial \theta_{2}^{\prime} \partial  \eta} &;~i=4\\
			-\frac{\partial ^2 \log(L(\Lambda))}{\partial \eta^{2}}&;~i=5,
			\end{cases},~V_{i4}=\begin{cases}
			-\frac{\partial ^2 \log(L(\Lambda))}{\partial \theta_{i} \partial \theta_{2}^{\prime} } &;~i=1,2\\
			-\frac{\partial ^2 \log(L(\Lambda))}{\partial \theta_{1}^{\prime} \partial  \theta_{2}^{\prime}} &;~i=3,
		\end{cases}
	\end{equation*}
	and $V_{12}=-\frac{\partial ^2 \log(L(\Lambda))}{\partial \theta_{1} \partial \theta_{2}}$, $V_{i3}=-\frac{\partial ^2 \log(L(\Lambda))}{\partial \theta_{i} \partial \theta_{1}^{\prime} };~i=1,2$.
	
\subsection{Bayesian Inference} \label{secBay}
	The Bayesian approach allows us to use the prior information about the parameters along with the observed data. The posterior distribution is given by,
	\begin{equation}\label{posterior}
		\pi (\Lambda|Y_1,Y_2) \propto L(Y_1,Y_2|\Lambda) \pi (\theta_1) \pi (\theta_2) \pi (\theta_1^{\prime}) \pi (\theta_2^{\prime}) \pi (\eta),
	\end{equation}
	where $\pi (\theta_1),\pi (\theta_2),\pi (\theta_1^{\prime}),\pi (\theta_2^{\prime})$ and $\pi (\eta)$ are the prior distributions of the parameters $\theta_1,\theta_2,\theta_1^{\prime},\theta_2^{\prime}$ and $\eta$ respectively and $L(Y_1,Y_2|\Lambda)$ is the likelihood function.
	\par The conditional posterior distributions does not correspond to known distributions and hence, Gibbs sampling cannot be performed to generate posterior samples. This complexity is handled using the Metropolis-Hastings (MH) Algorithm given below. This algorithm is based on a candidate generating (or proposal) density, $q(\Lambda, \nu)$, such that $\int q(\Lambda, \nu) d \nu=1$. The proposed candidate generating density could be symmetric or non-symmetric. We here consider Normal distribution as the proposed density for each of the parameters.
	\begin{itemize}
		\item \textit{Step 1}: Choose an arbitrary initial value $\Lambda_0$ and set iteration number $m=0$.
		\item \textit{Step 2}: Generate a candidate value $\Lambda^{*}$ from $q(\Lambda_m, \cdot)$ and a $u$ from $U(0,1)$.
		\item \textit{Step 3}: Calculate a probability of move $\alpha(\Lambda_m,\Lambda^{*})$ from $\Lambda$ to $\Lambda^{*}$ which is the ratio, $$\alpha(\Lambda_m,\Lambda^{*})=min \Bigg(\frac{\pi (\Lambda^{*}|Y_1,Y_2)q(\Lambda^{*},\Lambda)}{\pi (\Lambda|Y_1,Y_2)q(\Lambda,\Lambda^{*})},1 \Bigg).$$\\
		Since $q(\Lambda,v)$ is symmetric, $q(\Lambda^{*},\Lambda)=q(\Lambda,\Lambda^{*})$. If $u \leq \alpha(\Lambda_m,\Lambda^{*})$, accept the new candidate and set $\Lambda_{m+1}=\Lambda^{*}$, otherwise do not accept the new candidate and set $\Lambda_{m+1}=\Lambda_m$. Thus $\Lambda_{m+1}$ either takes the value $\Lambda^{*}$ or remains at the previous value $\Lambda_m$ at the $(m+1)^{th}$ iteration.
		\item \textit{Step 4}: Set $m=m+1$ and the above steps are repeated until convergence is attained.
 	\end{itemize}
	Thus we obtain a set of values which forms the sample from the posterior distribution. Convergence of these samples to the invariant target posterior distribution occurs only after the Markov chain has passed the transient stage and the effect of the fixed starting value $\Lambda_0$ has become negligible and can be ignored, and occurs under mild regularity conditions such as irreducibility and aperiodicity of the Markov chain. The posterior mode forms the Bayesian estimate, $\hat{\Lambda}$. Now to calculate the standard error of this estimate, we use the Bootstrap algorithm given by \cite{efron1986bootstrap} as follows.
	\subsubsection{Bootstrap algorithm}\label{bootstrap}
	\begin{itemize}
		\item \textit{Step 1}: Draw $B$ samples with replacement from the original data where each sample consists of $n$ observations. This forms the $B$ bootstrap samples.
		\item \textit{Step 2}: Generate the posterior sample corresponding to each bootstrap sample using the MH algorithm stated above.
		\item \textit{Step 3}: Calculate the mode of the posterior samples in Step 2 which forms the Bayesian estimate $\hat{\Lambda}_{b};~b=1,2,\dots,B$.
		\item \textit{Step 4}: Compute the average of the estimates calculated in Step 3 as $\hat{\Lambda}_{B}=\frac{1}{B} \sum_{b=1}^{B} \hat{\Lambda}_{b}$.
		\item \textit{Step 5}: Calculate the standard error of the Bayesian estimate as $\hat{\Lambda}_{SE}=\frac{1}{B-1} \sum_{b=1}^{B} (\hat{\Lambda}_{b}-\hat{\Lambda}_{B})^{2}$.
	\end{itemize}

	\section{Interval Estimation}\label{interval}
	In this section, we consider the asymptotic confidence interval and Bayesian credible intervals for the parameters in $\Lambda$ of the \textit{DPRH} model.
	
	\subsection{Asymptotic confidence intervals}\label{confidence}
	To find the $100(1-\alpha)\%$ confidence intervals of the parameters in the \textit{DPRH} model, we estimate the inverse of the observed Fisher information matrix given by $\hat{\Sigma}$ as in \eqref{covmatrix}. The confidence intervals for $\theta_{1},\theta_{2},\theta_{1}^{\prime},\theta_{2}^{\prime}$ and $\eta$ are then determined respectively as,
	\begin{equation*}
		\hat{\theta_{1}} \pm z_{\alpha/2}\sqrt{\text{var}(\hat{\theta_{1}})},\hat{\theta_{2}} \pm z_{\alpha/2}\sqrt{\text{var}(\hat{\theta_{2}})},\hat{\theta_{1}^{\prime}} \pm z_{\alpha/2}\sqrt{\text{var}(\hat{\theta_{1}^{\prime}})},\hat{\theta_{2}^{\prime}} \pm z_{\alpha/2}\sqrt{\text{var}(\hat{\theta_{2}^{\prime}})}~\text{and}~\hat{\eta} \pm z_{\alpha/2}\sqrt{\text{var}(\hat{\eta})},
	\end{equation*}
	where $z_{\alpha/2}$ is the upper $\alpha/2$$^{th}$ percentile of the standard normal distribution.
	
	\subsection{Bayesian credible intervals}\label{BCI}
	Let $\pi (\lambda|Y_1,Y_2)$ and $\Pi (\lambda|Y_1,Y_2)$ denote the marginal posterior density function and marginal posterior CDF of $\lambda$, respectively, where $\lambda \in \Lambda$. The $100(1-\alpha)\%$ Bayesian credible intervals for $\lambda$ is given by (\cite{chen2000computing}), 
	\begin{equation*}
		(\lambda^{(\alpha/2)},\lambda^{(1-\alpha/2)}),
	\end{equation*}
	which is also a high posterior density (HPD) interval if $\pi (\lambda|Y_1,Y_2)$ is symmetric and unimodal where $\Pi(\lambda^{(\alpha/2)}|Y_1,Y_2)=\alpha/2$ and $\Pi(\lambda^{(1-\alpha/2)}|Y_1,Y_2)=1-\alpha/2$. 
	
	\section{Simulation study} \label{simulation}
	In this section, we do an empirical study. For the Inverse Weibull baseline given by
	$F_{0}(y)=e^{-y^{-\alpha}};~~y>0,~\alpha>0$, with $\theta_{1}=\theta_{2}=\theta$, the \textit{DPRH} model is specified as, 
	\begin{align}\label{iwcdf2}
		\begin{split}
			F_{Y_{1}, Y_{2}}\left(y_{1}, y_{2}\right)&=\left\{\begin{array}{ll}{[F_{0}(y_2)]^{2\theta} F_{Y_1} \Big[\frac{1}{y_{1}^{-\alpha}-y_{2}^{-\alpha}} \Big]} & {\text {for}~~ y_{1}<y_{2}} \\ {[F_{0}(y_1)]^{2\theta} F_{Y_2} \Big[\frac{1}{y_{2}^{-\alpha}-y_{1}^{-\alpha}} \Big]} & {\text {for}~~ y_{1}>y_{2}}\end{array}\right. \\
			&=\left\{\begin{array}{ll}{e^{-2 \theta y_2^{-\alpha}} \Big[ \Big(\frac{\theta-\theta_1^{\prime}}{2\theta-\theta_1^{\prime}}\Big) e^{-2\theta (y_1^{-\alpha}-y_2^{-\alpha})} + \Big( \frac{\theta}{2\theta-\theta_{1}^{\prime}}\Big) e^{-\theta_{1}^{\prime}(y_1^{-\alpha}-y_2^{-\alpha})} \Big]} & {\text {for}~~ y_{1}<y_{2}} \\ {e^{-2 \theta y_1^{-\alpha}} \Big[ \Big(\frac{\theta-\theta_2^{\prime}}{2\theta-\theta_2^{\prime}}\Big) e^{-2\theta (y_2^{-\alpha}-y_1^{-\alpha})} + \Big( \frac{\theta}{2\theta-\theta_{2}^{\prime}}\Big) e^{-\theta_{2}^{\prime}(y_2^{-\alpha}-y_1^{-\alpha})} \Big]} & {\text {for}~~ y_{1}>y_{2}}\end{array}\right.
		\end{split}
	\end{align}
	The following algorithm can be used to generate a sample of size $n$ from the distribution in \eqref{iwcdf2} where the censoring scheme is introduced through the framework developed by \cite{wan2017simulating}.
	
	\subsection*{\normalfont Algorithm:}
	\begin{itemize}
		\item For each $k=1,2,\ldots,n$, generate  five independent Uniform $(0,1)$ random variables $U_{ki},~i=1,2,3,4,5$. 
		\item For a prefixed censoring percentage $p$ in the population, generate the censoring times $c_{1k}=z_{1}U_{k1}$ and $c_{2k}=z_{2}U_{k2}$ for $k=1,2,\ldots,n$, where $z_{1}$ and $z_{2}$ are derived by solving $P[0 \leq Y_{j} \leq c_{jk}, 0 \leq c_{jk} \leq z_{j}]=p;~j=1,2$.
		\item  If $U_{k3} \geq \frac{1}{2}$, set $t_{1k}=[-\frac{1}{2\theta}\text{log} (U_{k4})]^{-\frac{1}{\alpha}}$ and $t_{2k}=(t_{1k}^{-\alpha}+z_{k}^{-\alpha})^{-\frac{1}{\alpha}}$ where $z_{k}$ is the solution of $F_{Y_2}(z_k)=U_{k5}$.
		\item If $U_{k3} < \frac{1}{2}$, set $t_{2k}=[-\frac{1}{2\theta}\text{log} (U_{k4})]^{-\frac{1}{\alpha}}$ and $t_{1k}=(t_{2k}^{-\alpha}+z_{k}^{-\alpha})^{-\frac{1}{\alpha}}$ where $z_{k}$ is the solution of $F_{Y_1}(z_{k})=U_{k5}$.
		\item Take $y_{1k}=\text{Max}\{t_{1k},c_{1k}\}$ and $y_{2k}=\text{Max}\{t_{2k},c_{2k}\}$.
	\end{itemize}
	This gives us a bivariate left-censored sample of size $n$ from the \textit{DPRH} model in \eqref{iwcdf2}.
	\par We present the results for different sample sizes based on $r=500$ iterations to see how the MLE and Bayes estimators work in practice. The average bias across the $r$ samples and the mean square error (MSE) were computed as
	\begin{equation*}
		\text{Bias}=\frac{1}{r}\sum_{i=1}^{r}(\hat{\lambda_{i}}-\lambda)~~\text{and}~~
		\text{MSE}=\frac{1}{r}\sum_{i=1}^{r}(\hat{\lambda_{i}}-\lambda)^2,
	\end{equation*}
	where $\hat{\lambda_{i}}$ is the estimate of $\lambda$ in the $i^{th}$ iteration, for $\lambda \in \Lambda$. We have three different cases as given below. 
	\subsection{Case 1: MLE when $\theta$ known} \label{knowntheta}
	We have $\Lambda=(\theta,\theta,\theta_1^{\prime},\theta_2^{\prime},\alpha)$ as the vector of parameters where $\theta$ is known. Here the estimate of the dispersion matrix is given by
	\begin{equation*}
		\hat{\Sigma}=\begin{bmatrix}
			\text{var}(\hat{\theta_{1}^{\prime}}) & \text{cov}(\hat{\theta_{1}^{\prime}},\hat{\theta_{2}^{\prime}}) & \text{cov}(\hat{\theta_{1}^{\prime}},\hat{\eta}) \\& \text{var}(\hat{\theta_{2}^{\prime}}) & \text{cov}(\hat{\theta_{2}^{\prime}},\hat{\eta}) \\ & & \text{var}(\hat{\eta})
		\end{bmatrix}
	\end{equation*}
	where $\text{var}(\hat{\theta_{i}^{\prime}})=-\frac{\partial ^2 \log(L(\Lambda))}{\partial (\theta_{i}^{\prime})^{2}};~i=1,2$, $\text{var}(\hat{\eta})=-\frac{\partial ^2 \log(L(\Lambda))}{\partial \eta^{2}}$, $\text{cov}(\hat{\theta_{i}^{\prime}},\hat{\eta})=-\frac{\partial ^2 \log(L(\Lambda))}{\partial \theta_{i}^{\prime} \partial  \eta};~i=1,2$ and $\text{cov}(\hat{\theta_{1}^{\prime}},\hat{\theta_{2}^{\prime}})=-\frac{\partial ^2 \log(L(\Lambda))}{\partial \theta_{1}^{\prime} \partial  \theta_{2}^{\prime}}$. Hence the confidence intervals for $\theta_{1}^{\prime}, \theta_{2}^{\prime}$ and $\eta$ are computed respectively as,
	\begin{equation*}
		\hat{\theta_{1}^{\prime}} \pm z_{\alpha/2}\sqrt{\text{var}(\hat{\theta_{1}^{\prime}})},\hat{\theta_{2}^{\prime}} \pm z_{\alpha/2}\sqrt{\text{var}(\hat{\theta_{2}^{\prime}})}~\text{and}~\hat{\eta} \pm z_{\alpha/2}\sqrt{\text{var}(\hat{\eta})},
	\end{equation*}
	where $z_{\alpha/2}$ is the upper $\alpha/2$$^{th}$ percentile of the standard normal distribution. The results are reported in Table \ref{tabmle1}.
	\subsection{Case 2: MLE when $\theta$ unknown} \label{unknowntheta}
	We have $\Lambda=(\theta,\theta,\theta_1^{\prime},\theta_2^{\prime},\alpha)$ as the vector of parameters. The estimated dispersion matrix is
	\begin{equation*}
		\hat{\Sigma}=\begin{bmatrix}
		 \text{var}(\hat{\theta}) & \text{cov}(\hat{\theta},\hat{\theta_{1}^{\prime}}) & \text{cov}(\hat{\theta},\hat{\theta_{2}^{\prime}}) & \text{cov}(\hat{\theta},\hat{\eta}) \\ & \text{var}(\hat{\theta_{1}^{\prime}}) & \text{cov}(\hat{\theta_{1}^{\prime}},\hat{\theta_{2}^{\prime}}) & \text{cov}(\hat{\theta_{1}^{\prime}},\hat{\eta}) \\ & & \text{var}(\hat{\theta_{2}^{\prime}}) & \text{cov}(\hat{\theta_{2}^{\prime}},\hat{\eta}) \\ & & & \text{var}(\hat{\eta})
		\end{bmatrix}
	\end{equation*}
	where $\text{var}(\hat{\theta})=-\frac{\partial ^2 \log(L(\Lambda))}{\partial (\theta)^{2}}$, $\text{var}(\hat{\theta_{i}^{\prime}})=-\frac{\partial ^2 \log(L(\Lambda))}{\partial (\theta_{i}^{\prime})^{2}};~i=1,2$, $\text{var}(\hat{\eta})=-\frac{\partial ^2 \log(L(\Lambda))}{\partial \eta^{2}}$, $\text{cov}(\hat{\theta},\hat{\theta_{i}^{\prime}})=-\frac{\partial ^2 \log(L(\Lambda))}{\partial \theta \partial  \theta_{i}^{\prime}};~i=1,2$, $\text{cov}(\hat{\theta_{i}^{\prime}},\hat{\eta})=-\frac{\partial ^2 \log(L(\Lambda))}{\partial \theta_{i}^{\prime} \partial  \eta};~i=1,2$ and $\text{cov}(\hat{\theta_{1}^{\prime}},\hat{\theta_{2}^{\prime}})=-\frac{\partial ^2 \log(L(\Lambda))}{\partial \theta_{1}^{\prime} \partial  \theta_{2}^{\prime}}$. The confidence intervals for $\theta,\theta_{1}^{\prime},\theta_{2}^{\prime}$ and $\eta$ are determined respectively as,
	\begin{equation*}
		\hat{\theta} \pm z_{\alpha/2}\sqrt{\text{var}(\hat{\theta})},\hat{\theta_{1}^{\prime}} \pm z_{\alpha/2}\sqrt{\text{var}(\hat{\theta_{1}^{\prime}})},\hat{\theta_{2}^{\prime}} \pm z_{\alpha/2}\sqrt{\text{var}(\hat{\theta_{2}^{\prime}})}~\text{and}~\hat{\eta} \pm z_{\alpha/2}\sqrt{\text{var}(\hat{\eta})},
	\end{equation*}
	where $z_{\alpha/2}$ is the upper $\alpha/2$$^{th}$ percentile of the standard normal distribution. The results are reported in Table \ref{tabmle2}.
	\par The MLE of the parameters are obtained by optimising the log-likelihood function in \eqref{likelihood} using the $optim$ function in $R$. The dispersion matrix $\Sigma$ is estimated as $\hat{\Sigma}$ and the $95\%$ confidence interval for each of the parameters in each iteration are computed corresponding to each cases in Sections \ref{knowntheta} and \ref{unknowntheta}. Thus, the coverage probability can be determined by computing the proportion of samples for which the population parameter is contained in the confidence interval. It is observed that the model works well with large sample sizes in particular.
	\subsection{Case 3: Bayesian Inference when $\theta$ unknown}
	Since all the parameters are positive, we assume they have a Gamma$(m,p)$ prior for $m > 0$ and $p > 0$, that is,
	\begin{equation*}
		\pi(\lambda | m,p)=\left\{\begin{array}{ll}{\frac{m^p}{\Gamma (p)} e^{-m \lambda} \lambda^{p-1}} & {;~ \lambda > 0} \\ {0} & {;~ \text{otherwise}}\end{array}\right.,
	\end{equation*}
	where $\lambda \in \Lambda$. The prior parameters $m$ and $p$ are chosen such that the mean $\frac{m}{p}$ is the maximum likelihood estimates obtained in Table \ref{tabmle2} and variance $\frac{m}{p^2}$ is $1.2$. The results are reported in Table \ref{tabbay1}.
	\par We also carried out the Bayesian analysis with a Normal$(\mu,\sigma)$ prior to rule out prior sensitivity, where for $-\infty < \mu < \infty,~\sigma > 0$ and  $\lambda \in \Lambda$,
	\begin{equation*}
		\pi(\lambda | \mu,\sigma)=\left\{\begin{array}{ll}{\frac{1}{\sigma \sqrt{2 \pi}}e^{-\frac{1}{2 \sigma ^2}(\lambda-\mu)^{2}}} & {;~ -\infty < \lambda < \infty} \\ {0} & {;~ \text{otherwise}}\end{array}\right..
	\end{equation*}
	The prior parameters $\mu$ and $\sigma$ are chosen such that $\mu$ is the MLE and $\sigma$ is $0.1$. The results are reported in Table \ref{tabbay2}.
	\par The posterior sample is obtained using the \textit{MHadaptive} package in R software. The Bayesian analysis seem to work well and the posterior mode is reported as the Bayesian estimate as we obtain unimodal posteriors of the data.
	\begin{table}
		\centering
		\caption{Maximum likelihood estimates, Bias, MSEs and Coverage Probabilities of the parameters based on $500$ iterations with Inverse Weibull distribution baseline and $\Lambda=(1.3,1.3,1.5,1.6,1.2)$ when $\theta_{1}=\theta_{2}=\theta$ is known \label{tabmle1}}
		\begin{tabular}{cllll}
			\hline
			\multicolumn{1}{l}{\textbf{Sample size}} &                  & $\theta_1^{\prime}$ & $\theta_2^{\prime}$ & $\alpha$ \\ \hline
			\multirow{4}{*}{n=30}           & Estimates         & 1.6119           & 1.7433             & 1.2498   \\  
			& Bias               & 0.1119             & 0.1433             & 0.0498   \\  
			& MSE               & 1.1525              & 1.2879              & 0.0304   \\ 
			& Cov. Probability   & 0.9235              & 0.8815              & 0.9559   \\ \hline
			\multirow{4}{*}{n=100}          & Estimates         & 1.4117             & 1.4286              & 1.2229   \\  
			& Bias              & -0.0883             & -0.1714              & 0.0229   \\  
			& MSE                & 0.1826              & 0.2059             & 0.0081 \\  
			& Cov. Probability   & 0.9000              & 0.8620             & 0.9560   \\ \hline
		\end{tabular}
	\end{table}
	\begin{table}
		\centering
		\caption{Maximum likelihood estimates, Bias, MSEs and Coverage Probabilities of the parameters based on $500$ iterations with Inverse Weibull distribution baseline and $\Lambda=(1.5,1.5,1.7,1.8,1.3)$ \label{tabmle2}}
		\begin{tabular}{clllll}
			\hline
			\multicolumn{1}{l}{\textbf{Sample size}} &                  & $\theta$ & $\theta_1^{\prime}$ & $\theta_2^{\prime}$ & $\alpha$ \\ \hline
			\multirow{4}{*}{n=30}           & Estimates        & 1.7651   & 1.8684              & 1.9875             & 1.4025   \\  
			& Bias             & 0.2651   & 0.1684              & 0.1875              & 0.1025   \\  
			& MSE              & 0.5592   & 1.4362              & 1.3560              & 0.0723   \\ 
			& Cov. Probability & 0.9540   & 0.9357              & 0.9178              & 0.9660   \\ \hline
			\multirow{4}{*}{n=100}          & Estimates        & 1.6113   & 1.6337              & 1.6891              & 1.3484   \\  
			& Bias             & 0.1113   & -0.0663              & -0.1109              & 0.0484   \\  
			& MSE              & 0.1248   & 0.2121              & 0.2636             & 0.0197  \\  
			& Cov. Probability & 0.9560   & 0.9280              & 0.8700              & 0.9400   \\ \hline
		\end{tabular}
	\end{table}
	
	\begin{table}
		\centering
		\caption{Bayesian estimates, Bias and MSEs of the parameters based on $500$ iterations with Inverse Weibull distribution baseline, $Gamma$ prior and $\Lambda=(1.5,1.5,1.7,1.8,1.3)$ \label{tabbay1}}
		\begin{tabular}{clllll}
			\hline
			\multicolumn{1}{l}{\textbf{Sample size}} &           & $\theta$                   & $\theta_1^{\prime}$ & $\theta_2^{\prime}$ & $\alpha$ \\ \hline
			\multirow{3}{*}{n=30}                            & Estimates & 1.4682                     & 1.4734              & 1.5304              & 1.2786  \\
			& Bias      & -0.0318                    & -0.2266             & -0.2696             & 0.0214   \\
			& MSE       & 0.2106                  & 0.4244              & 0.4227             & 0.0477   \\ \hline
			\multirow{3}{*}{n=100}                           & Estimates &    1.4950                 &    1.5094         &      1.5103       &  1.2968  \\
			& Bias      &  -0.0050                    &  -0.1906         &        -0.2897       & -0.0032  \\
			& MSE       &        0.0792              &     0.2116          &      0.2871        &  0.0181 \\ \hline
		\end{tabular}
	\end{table}

	\begin{table}
		\centering
		\caption{Bayesian estimates, Bias and MSEs of the parameters based on $500$ iterations with Inverse Weibull distribution baseline, $Normal$ prior and $\Lambda=(1.5,1.5,1.7,1.8,1.3)$ \label{tabbay2}}
		\begin{tabular}{clllll}
			\hline
			\multicolumn{1}{l}{\textbf{Sample size}} &           & $\theta$ & $\theta_1^{\prime}$ & $\theta_2^{\prime}$ & $\alpha$ \\ \hline
			\multirow{3}{*}{n=30}                    & Estimates &   1.7484      &        1.8584          &    1.9737                &  1.3874    \\
			& Bias      &    0.2484  &   0.1584                &        0.1737           &     0.0874  \\
			& MSE       &   0.0658     &    0.0287               &        0.0339           &   0.0122      \\ \hline
			\multirow{3}{*}{n=100}                   & Estimates & 1.5940  & 1.6243   &  1.6825 & 1.3353  \\
			& Bias      & 0.0940   &  -0.0757              &   -0.1175     &   0.0353   \\
			& MSE       &   0.0131   &      0.0095         &    0.0177    &  0.0054 \\ \hline
		\end{tabular}
	\end{table}
	
\section{Application}\label{realdata}
\subsection*{\textbf{Data Description}}
	The Australian twin data (\cite{duffy1990appendectomy}) consists of $3808$ pair of twins who responded to a questionnaire prepared by the Australian Twin Registry in $1982$. The data has information on the age, sex, zygosity of the twins and whether they had undergone appendectomy. Appendectomy is a low risk emergency surgery to remove an infected appendix in humans. Intravenous medicines are also effective in curing appendicitis. Once a subject undergoes appendectomy, it ensures a risk free time for the rest of the subject's life. We here are interested in the distribution of the residual life of twins after appendectomy, referred to as risk free time henceforth.
	\par The data has six zygote categories of which we consider the second category which consists of monozygotic (identical) male-male pair as the dependence is more significant in them than the dizygotic (non-identical) twins (\cite{fortuna2010twin}) and the genetic effect can be advocated into the model through this dependence existing between the twins. The simultaneous failures are discarded. There are $157$ pairs of twins with an indicator variable for each twin in a pair denoted by $ c_1 $ and $ c_2 $ representing the presence or absence of appendectomy among the twins. Let $(T_1,T_2)$ denote the onset age at appendectomy or the censoring age of the twins. Then $(Y_1,Y_2)=(b-T_1,b-T_2)$ is the risk free time of the twins assuming the twins live up to an age of $b$ years. We choose $b$ as $80$. Observe that $(Y_{1},Y_{2})$ corresponds to left-censored observations.
\subsection*{\textbf{Data Analysis}}
	Since the data pertains to twins, we assume $\theta_{i}=\theta$ and $\theta_{i}^{\prime}=\theta^{\prime};~i=1,2$. The \textit{DPRH} model with Generalized Rayleigh, Exponentiated Gumbel and Generalized Exponential baselines is used to analyse the data. The parameters are estimated for these baseline distributions and is summarised in Table \ref{tabdmle}. The standard errors of the estimates are obtained from the diagonal elements of the inverse of the observed Fisher Information matrix. We also calculated the $95\%$ confidence intervals for the parameters using the results in Section \ref{confidence}. The Akaike information criterion (AIC) value for each model is also given for comparing the models which is calculated as $\text{AIC}=-2 \log L+2p$, where $p$ is the number of parameters in the model. From the AIC values, we conclude that the bivariate proportional reversed hazard model with baseline distribution as Generalized Rayleigh is a better model for the Australian Twin dataset.
	\par
\begin{figure}	
	 	\centering
	 	\begin{minipage}{0.45\textwidth}
	 		\includegraphics[width=0.9\textwidth]{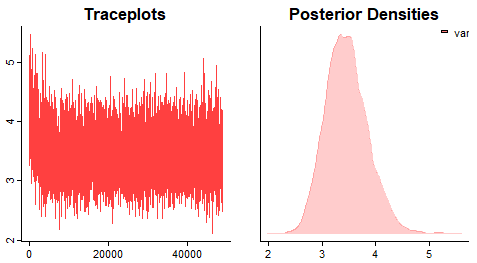} 
	 		\subcaption{}
	 	\end{minipage}\hfill
	 	\begin{minipage}{0.45\textwidth}
	 		\centering
	 		\includegraphics[width=0.9\textwidth]{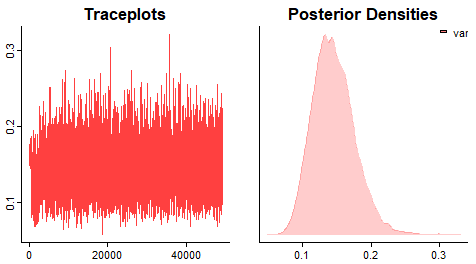} 
	 		\subcaption{}
	 	\end{minipage}
 		\begin{minipage}{0.45\textwidth}
 			\includegraphics[width=0.9\textwidth]{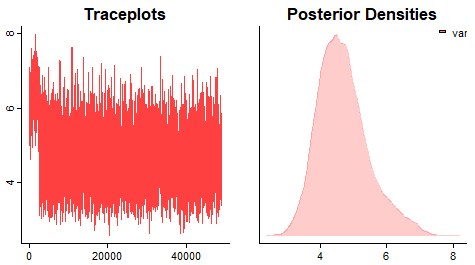} 
 			\subcaption{}
 		\end{minipage}\hfill
 		\begin{minipage}{0.45\textwidth}
 			\centering
 			\includegraphics[width=0.9\textwidth]{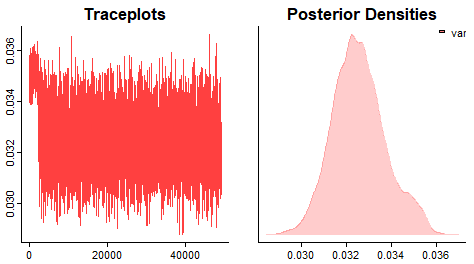} 
 			\subcaption{}
 		\end{minipage}	
	 	\caption{Trace plots and posterior density plots of (a) $\theta$ (b) $\theta^{\prime}$ (c) $\alpha$ and (d) $\lambda$ with the baseline distribution as Generalized Rayleigh}
	 	\label{figplots}
	 \end{figure}

 The Bayesian estimates with a Generalized Rayleigh baseline and Gamma priors are computed. The samples from the posterior distribution is generated using the MH algorithm given in Section \ref{secBay}. We take the posterior mode as the Bayesian estimate for each of the parameters which are obtained as $\hat{\theta}_{B}=3.1430$, $\hat{\theta^{\prime}}_{B}=0.1279$, $\hat{\alpha}_{B}=4.4422$ and $\hat{\lambda}_{B}=0.0322$. The trace plots and posterior density plots are given in Figure \ref{figplots}. Since the posterior densities are symmetric, we calculate the $95\%$ Bayesian credible intervals for the parameters, which is also an HPD interval, using the results in Section \ref{BCI}. The standard error of the Bayesian estimates are obtained using the Bootstrap algorithm given in Section \ref{bootstrap} and is reported in Table \ref{tabdbay}  along with the corresponding Bayesian credible intervals and posterior mean. A likelihood ratio test to test,
	 $H_0: \theta=\theta^{\prime}~\textit{vs.}~H_1: \theta \neq \theta^{\prime}$ yielded a high chi square value, thereby confirming a dependence between a pair of twins.
	 \par We calculated the probabilities, $P[Y_{1}\leq y_{1}| Y_{2}=y_{2}]$ for $y_{1}<y_{2}$ and $P[Y_{2} \leq y_{2} | Y_{1}=y_{1}]$ for $y_{1}>y_{2}$. Observe that $P[Y_{1}\leq y_{1}| Y_{2}=y_{2}]=P[T_{1}\geq t_{1}| T_{2}=t_{2}]$ for $t_{1}>t_{2}$ is the probability of twin 1 to undergo appendectomy at some time after $t_{1}$ given that his co-twin has already undergone appendectomy at $t_{2}$. With a similar interpretation, we have $P[Y_{2}\leq y_{2}| Y_{1}=y_{1}]=P[T_{2}\geq t_{2}| T_{1}=t_{1}]$ for $t_{2}>t_{1}$. We computed these probabilities of potential appendectomy using our fitted model for all the $157$ cases and checked the consistency with the observed data (a few cases are given in Table \ref{tabvalidation}). We observed that co-twins for whom the appendectomy has not occurred showed a high probability whereas in the case where appendectomy had already been done for both twins, the model gave relatively low probabilities for a potential appendectomy. This was obtained for 73\% of the data points. This is proposed as a validation for our fitted model. Hence, we have proposed the \textit{DPRH} model with Generalized Rayleigh distribution as the baseline distribution for modelling the Australian twin dataset with parameter values as $\theta=5.3398,\theta^{\prime}=0.1453,\alpha=4.7711$ and $\lambda=0.0343$. 
 
 	\begin{table}[]
 		\centering
 		\caption{Maximum Likelihood Estimates of the parameters for various baseline distributions for the Australian twin dataset\label{tabdmle}}
 		\begin{tabular}{lcccccl}
 			\hline
 			\multirow{2}{*}{\begin{tabular}[c]{@{}l@{}}\textbf{Baseline}\\ \textbf{Distribution}\end{tabular}}          & \multicolumn{1}{c}{\multirow{2}{*}{\textbf{Parameters}}} & \multirow{2}{*}{\textbf{Estimate}} & \multirow{2}{*}{\begin{tabular}[c]{@{}c@{}}\textbf{Standard}\\ \textbf{Error}\end{tabular}} & \multirow{2}{*}{\textbf{LCL}} & \multirow{2}{*}{\textbf{UCL}} & \multicolumn{1}{c}{\multirow{2}{*}{\textbf{AIC}}} \\ 
 			& \multicolumn{1}{c}{}                            &                           &                                                                           &                      &                      & \multicolumn{1}{c}{}                     \\ \hline
 			\multirow{4}{*}{\begin{tabular}[c]{@{}l@{}}Generalized\\ Rayleigh\end{tabular}}           & $\theta$                                        & 5.3398                    &      1.1128                                                               &      3.1589        &    7.5208          & \multirow{4}{*}{1759.80}                 \\
 			& $\theta^{\prime}$                               & 0.1453                    &    0.0066                                                               &     0.1324        &   0.1582           &                                          \\ 
 			& $\alpha$                                        & 4.7711                    &         0.3210                                                              &     4.1420         &     5.4003         &                                          \\
 			& $\lambda$                                       & 0.0343                   &      0.0013                                                            &       0.0318        &      0.0368        &                                          \\ \hline
 			\multirow{3}{*}{\begin{tabular}[c]{@{}l@{}}Exponentiated\\ Gumbel\end{tabular}}           & $\theta$                                        &    33.2154                       &   7.1792                                                                       &       19.1443              &    47.2866                 & \multirow{3}{*}{1852.45}                 \\
 			& $\theta^{\prime}$                               &        2.9245                   &       0.5649                                                                   &      1.8173                &       4.0317              &                                          \\
 			& $\lambda$                                       &         0.0715                  &      0.0038                                                                     &      0.0641               &       0.0789  	                 &                                     \\ \hline
 			\multirow{3}{*}{\begin{tabular}[c]{@{}l@{}}Generalized\\ Exponential\end{tabular}}        & $\theta$                                        &                      34.1203     &        7.8001                                                                   &       18.8322               &       49.4084               & \multirow{3}{*}{1867.77}                 \\
 			& $\theta^{\prime}$                               &         2.4048                  &       0.4746                                                                    &     1.4745                 &         3.3350               &                                        \\
 			& $\lambda$                                       &       0.0722                    &    0.0040                                                                       &       0.0644               &         0.0800                &                                       \\ \hline
 		\end{tabular}
 	\end{table}
 	
\begin{table}[]
	\centering
	\caption{$95\%$ Bayesian credible intervals, Posterior mean and Posterior mode of the estimates \label{tabdbay}}
	\begin{tabular}{ccccc}
		\hline
		\multirow{2}{*}{\textbf{Parameter}} & \multirow{2}{*}{\textbf{0.025}} & \multirow{2}{*}{\textbf{0.975}} & \multirow{2}{*}{\textbf{\begin{tabular}[c]{@{}c@{}}Posterior Mean\\ (Standard Error)\end{tabular}}} & \multirow{2}{*}{\textbf{\begin{tabular}[c]{@{}c@{}}Posterior Mode\\ (Standard Error)\end{tabular}}} \\ 
		&                                 &                                 &                                          &                                                                                                     \\ \hline
		$\theta$                            &            2.7601             &    4.2815                   &                3.4590 (0.0281)                 & 3.1430 (0.4634)                                                                                     \\ 
		$\theta^{\prime}$                   &     0.0927                   &     0.2101                     &            0.1450 (0.0015)                     & 0.1279 (0.0159)                                                                                     \\ 
		$\alpha$      &     3.4032                     &          6.4969                 &      4.6746 (0.0452)                            &  4.4422(0.4664)    \\ 
		$\lambda$      &      0.0304                    &        0.0352                 &     0.0326  (6.4 $\times$ $10^{-5}$)                             & 0.0322 (0.0008)    \\ \hline
	\end{tabular}
\end{table}

\begin{table}[]
	\centering
	\caption{Calculated probabilities for validation\label{tabvalidation}}
	\begin{tabular}{ccccc}
		\hline
		\multirow{2}{*}{\textbf{$t_1$}} & \multirow{2}{*}{\textbf{$c_1$}} & \multirow{2}{*}{\textbf{$t_2$}} & \multirow{2}{*}{\textbf{$c_2$}} & \multirow{2}{*}{\textbf{\begin{tabular}[c]{@{}c@{}}Probability of\\ potential appendectomy\end{tabular}}} \\ 
		&                                 &                                 &                                          &                                                                                                     \\ \hline
		36             & 0              & 11             & 1              & 0.8981               \\
		15             & 1              & 27             & 0              & 0.9405               \\
		33             & 0              & 8              & 1              & 0.9181               \\
		18             & 1              & 45             & 0              & 0.8070               \\
		27             & 1              & 31             & 1              & 0.8769               \\
		11             & 1              & 26             & 0              & 0.9475               \\
		44             & 0              & 15             & 1              & 0.8229               \\
		40             & 1              & 52             & 1              & 0.3206               \\
		45             & 1              & 56             & 1              & 0.0797               \\
		46             & 0              & 25             & 1              & 0.7717              \\ \hline
	\end{tabular}
\end{table}

\section{Conclusion} \label{conclusion}
We have proposed a class of bivariate models which can be used to model a two-component load sharing system when there are left-censored observations. The proposed model uses the concept of proportional reversed hazard rate for modelling such systems. It has explicit forms for the probability density function and cumulative distribution functions. Various properties of the model have been discussed. Also, different inferential procedures such as the maximum likelihood estimation and Bayesian estimation are discussed. We also provide the asymptotic confidence intervals and Bayesian credible intervals for the parameters in the model. Simulation study and real data analysis are given to illustrate the proposed model.
\par The usefulness of the model was illustrated by analysing the Australian twin data (\cite{duffy1990appendectomy}). The likelihood ratio test confirmed the dependence.


\section{Appendix}
\begin{app}\label{app1}
	The likelihood function given the data can be obtained as follows:
	\begin{equation*} \label{censlike}
		L\left(\Lambda \right)=\prod_{k \in I_1} f_{1k} \prod_{k \in I_2} f_{2k} \prod_{k \in I_3} f_{3k} \prod_{k \in I_4} f_{4k} \prod_{k \in I_5} f_{5k} \prod_{k \in I_6} f_{6k} \prod_{k \in I_7} f_{7k} \prod_{k \in I_8} f_{8k},
	\end{equation*}
	where,
	\begin{align*}
		f_{1k}&=\frac{\partial 	^{2}F_{y_{1k}>y_{2k}}(y_{1k},y_{2k})}{\partial y_{1k} \partial y_{2k}}=\theta_{1} \theta_{2}^{\prime} f_{0}(y_{1k}) f_{0}(y_{2k}) [F_{0}(y_{1k})]^{\theta_1 +\theta_2-\theta_2^{\prime}-1} [F_{0}(y_{2k})]^{\theta_2^{\prime}-1}\\
		f_{2k}&=\frac{\partial ^{2}F_{y_{1k}<y_{2k}}(y_{1k},y_{2k})}{\partial y_{1k} \partial y_{2k}}=\theta_{1}^{\prime} \theta_{2} f_{0}(y_{1k}) f_{0}(y_{2k}) [F_{0}(y_{1k})]^{\theta_1^{\prime}-1} [F_{0}(y_{2k})]^{\theta_1 +\theta_2-\theta_1^{\prime}-1}\\
		f_{3k}&=\frac{\partial F_{y_{1k}>c_{2k}}(y_{1k},c_{2k})}{\partial y_{1k}}=\theta_{1} f_{0}(y_{1k}) [F_{0}(y_{1k})]^{\theta_1 +\theta_2-\theta_2^{\prime}-1} [F_{0}(y_{2k})]^{\theta_2^{\prime}}\\
		f_{4k}&=\frac{\partial F_{c_{1k}<y_{2k}}(c_{1k},y_{2k})}{\partial y_{2k}}=\theta_{2}  f_{0}(y_{2k}) [F_{0}(y_{1k})]^{\theta_1^{\prime}} [F_{0}(y_{2k})]^{\theta_1 +\theta_2-\theta_1^{\prime}-1}\\
		f_{5k}&=\frac{\partial F_{y_{1k}<c_{2k}}(y_{1k},c_{2k})}{\partial y_{1k}}=\frac{(\theta_{1}-\theta_{1}^{\prime})(\theta_{1}+\theta_{2})}{\theta_{1}+\theta_{2}-\theta_{1}^{\prime}} f_{0}(y_{1k}) [F_{0}(y_{1k})]^{\theta_1 +\theta_2-1}\\
		&  \,\,\,\,\,\,\,\,\,\,\,\,\,\,\,\,\,\,\,\,\,\,\,\,\,\,\,\,\,\,\,\,\,\,\,\,\,\,\,\,\,\,\,\,\,\,\,\,\,\,\,\,\,\,\,\,\,\,\,\,\,\,\,\,\,\,\,\,\,\,\,\,\,\,\,\,\,\,\,\, +\frac{\theta_{1}^{\prime}\theta_{2}}{\theta_{1}+\theta_{2}-\theta_{1}^{\prime}}f_{0}(y_{1k}) [F_{0}(y_{1k})]^{\theta_{1}-1} [F_{0}(y_{2k})]^{\theta_1 +\theta_2-\theta_{1}^{\prime}} \\
		f_{6k}&=\frac{\partial F_{c_{1k}>y_{2k}}(c_{1k},y_{2k})}{\partial y_{2k}}=\frac{(\theta_{2}-\theta_{2}^{\prime})(\theta_{1}+\theta_{2})}{\theta_{1}+\theta_{2}-\theta_{2}^{\prime}} f_{0}(y_{2k}) [F_{0}(y_{2k})]^{\theta_1 +\theta_2-1}\\
		& \,\,\,\,\,\,\,\,\,\,\,\,\,\,\,\,\,\,\,\,\,\,\,\,\,\,\,\,\,\,\,\,\,\,\,\,\,\,\,\,\,\,\,\,\,\,\,\,\,\,\,\,\,\,\,\,\,\,\,\,\,\,\,\,\,\,\,\,\,\,\,\,\,\,\,\,\,\,\,\, +\frac{\theta_{1}\theta_{2}^{\prime}}{\theta_{1}+\theta_{2}-\theta_{2}^{\prime}}f_{0}(y_{2k}) [F_{0}(y_{1k})]^{\theta_1 +\theta_2-\theta_{2}^{\prime}} [F_{0}(y_{2k})]^{\theta_{2}-1} \\
		f_{7k}&=F_{c_{1k}>c_{2k}}(c_{1k},c_{2k})=\frac{\theta_{2}-\theta_{2}^{\prime}}{\theta_{1}+\theta_{2}-\theta_{2}^{\prime}} [F_{0}(y_{2k})]^{\theta_1 +\theta_2}+\frac{\theta_{1}}{\theta_{1}+\theta_{2}-\theta_{2}^{\prime}} [F_{0}(y_{1k})]^{\theta_1 +\theta_2-\theta_{2}^{\prime}} [F_{0}(y_{2k})]^{\theta_{2}^{\prime}}\\
		f_{8k}&=F_{c_{1k}<c_{2k}}(c_{1k},c_{2k})=\frac{\theta_{1}-\theta_{1}^{\prime}}{\theta_{1}+\theta_{2}-\theta_{1}^{\prime}} [F_{0}(y_{1k})]^{\theta_1 +\theta_2}+\frac{\theta_{2}}{\theta_{1}+\theta_{2}-\theta_{1}^{\prime}} [F_{0}(y_{1k})]^{\theta_{1}^{\prime}} [F_{0}(y_{2k})]^{\theta_1 +\theta_2-\theta_{1}^{\prime}}.
	\end{align*}
	Hence the log-likelihood function is given by,
	\begin{align}
		\nonumber \log L(\Lambda)=&\sum_{k \in I_1} \theta_{1} \theta_{2}^{\prime} f_{0}(y_{1k}) f_{0}(y_{2k}) [F_{0}(y_{1k})]^{\theta_1 +\theta_2-\theta_2^{\prime}-1} [F_{0}(y_{2k})]^{\theta_2^{\prime}-1}  \\ \nonumber 
		&+ \sum_{k \in I_2}\theta_{1}^{\prime} \theta_{2} f_{0}(y_{1k}) f_{0}(y_{2k}) [F_{0}(y_{1k})]^{\theta_1^{\prime}-1} [F_{0}(y_{2k})]^{\theta_1 +\theta_2-\theta_1^{\prime}-1} \\ 
		&+ \sum_{k \in I_3}\theta_{1} f_{0}(y_{1k}) [F_{0}(y_{1k})]^{\theta_1 +\theta_2-\theta_2^{\prime}-1} [F_{0}(y_{2k})]^{\theta_2^{\prime}} \label{likelihood} \\ \nonumber 
		&+ \sum_{k \in I_4}\theta_{2}  f_{0}(y_{2k}) [F_{0}(y_{1k})]^{\theta_1^{\prime}} [F_{0}(y_{2k})]^{\theta_1 +\theta_2-\theta_1^{\prime}-1} \\ \nonumber
		&+ \sum_{k \in I_5} \frac{(\theta_{1}-\theta_{1}^{\prime})(\theta_{1}+\theta_{2})}{\theta_{1}+\theta_{2}-\theta_{1}^{\prime}} f_{0}(y_{1k}) [F_{0}(y_{1k})]^{\theta_1 +\theta_2-1}
	\end{align}
	\begin{align*}
		&+ \sum_{k \in I_5} \frac{\theta_{1}^{\prime}\theta_{2}}{\theta_{1}+\theta_{2}-\theta_{1}^{\prime}}f_{0}(y_{1k}) [F_{0}(y_{1k})]^{\theta_{1}-1} [F_{0}(y_{2k})]^{\theta_1 +\theta_2-\theta_{1}^{\prime}} \\ \nonumber 
		& + \sum_{k \in I_6} \frac{(\theta_{2}-\theta_{2}^{\prime})(\theta_{1}+\theta_{2})}{\theta_{1}+\theta_{2}-\theta_{2}^{\prime}} f_{0}(y_{2k}) [F_{0}(y_{2k})]^{\theta_1 +\theta_2-1} \\ \nonumber 
		& + \sum_{k \in I_6} \frac{\theta_{1}\theta_{2}^{\prime}}{\theta_{1}+\theta_{2}-\theta_{2}^{\prime}}f_{0}(y_{2k}) [F_{0}(y_{1k})]^{\theta_1 +\theta_2-\theta_{2}^{\prime}} [F_{0}(y_{2k})]^{\theta_{2}-1} \\ \nonumber 
		& + \sum_{k \in I_7} \Big[ \frac{\theta_{2}-\theta_{2}^{\prime}}{\theta_{1}+\theta_{2}-\theta_{2}^{\prime}} [F_{0}(y_{2k})]^{\theta_1 +\theta_2}+\frac{\theta_{1}}{\theta_{1}+\theta_{2}-\theta_{2}^{\prime}} [F_{0}(y_{1k})]^{\theta_1 +\theta_2-\theta_{2}^{\prime}} [F_{0}(y_{2k})]^{\theta_{2}^{\prime}} \Big]\\ \nonumber 
		& + \sum_{k \in I_8} \Big[\frac{\theta_{1}-\theta_{1}^{\prime}}{\theta_{1}+\theta_{2}-\theta_{1}^{\prime}} [F_{0}(y_{1k})]^{\theta_1 +\theta_2}+\frac{\theta_{2}}{\theta_{1}+\theta_{2}-\theta_{1}^{\prime}} [F_{0}(y_{1k})]^{\theta_{1}^{\prime}} [F_{0}(y_{2k})]^{\theta_1 +\theta_2-\theta_{1}^{\prime}}\Big].
	\end{align*}
\end{app}


\bibliography{ref}

\begin{thebibliography}{}

\bibitem[Balakrishnan et~al., 2021]{balakrishnan2021comparisons}
Balakrishnan, N., Barmalzan, G., and Kosari, S. (2021).
\newblock Comparisons of parallel systems with components having proportional
  reversed hazard rates and starting devices.
\newblock {\em Mathematics}, 9(8):856.

\bibitem[Barlow et~al., 1963]{barlow1963properties}
Barlow, R.~E., Marshall, A.~W., Proschan, F., et~al. (1963).
\newblock Properties of probability distributions with monotone hazard rate.
\newblock {\em The Annals of Mathematical Statistics}, 34(2):375--389.

\bibitem[Bismi, 2005]{bismi2005bivariate}
Bismi, G. (2005).
\newblock Bivariate burr distributions.
\newblock {\em PhD Thesis, Cochin University of Science and Technology}.

\bibitem[Block et~al., 1998]{block1998reversed}
Block, H.~W., Savits, T.~H., and Singh, H. (1998).
\newblock The reversed hazard rate function.
\newblock {\em Probability in the Engineering and informational Sciences},
  12(1):69--90.

\bibitem[Chen et~al., 2000]{chen2000computing}
Chen, M.-H., Shao, Q.-M., and Ibrahim, J.~G. (2000).
\newblock Computing bayesian credible and hpd intervals.
\newblock In {\em Monte Carlo Methods in Bayesian Computation}, pages 213--235.
  Springer.

\bibitem[Dabrowska et~al., 1988]{dabrowska1988kaplan}
Dabrowska, D.~M. et~al. (1988).
\newblock Kaplan-meier estimate on the plane.
\newblock {\em Annals of Statistics}, 16(4):1475--1489.

\bibitem[Di~Crescenzo, 2000]{di2000some}
Di~Crescenzo, A. (2000).
\newblock Some results on the proportional reversed hazards model.
\newblock {\em Statistics \& probability letters}, 50(4):313--321.

\bibitem[Duffy et~al., 1990]{duffy1990appendectomy}
Duffy, D.~L., Martin, N.~G., and Mathews, J.~D. (1990).
\newblock Appendectomy in australian twins.
\newblock {\em American journal of human genetics}, 47(3):590.

\bibitem[Efron and Tibshirani, 1986]{efron1986bootstrap}
Efron, B. and Tibshirani, R. (1986).
\newblock Bootstrap methods for standard errors, confidence intervals, and
  other measures of statistical accuracy.
\newblock {\em Statistical science}, pages 54--75.

\bibitem[Fortuna et~al., 2010]{fortuna2010twin}
Fortuna, K., Goldner, I., and Knafo, A. (2010).
\newblock Twin relationships: A comparison across monozygotic twins, dizygotic
  twins, and nontwin siblings in early childhood.
\newblock {\em Family Science}, 1(3-4):205--211.

\bibitem[Ganser and Hewett, 2010]{ganser2010accurate}
Ganser, G.~H. and Hewett, P. (2010).
\newblock An accurate substitution method for analyzing censored data.
\newblock {\em Journal of occupational and environmental hygiene},
  7(4):233--244.

\bibitem[Gupta et~al., 1998]{gupta1998modeling}
Gupta, R.~C., Gupta, P.~L., and Gupta, R.~D. (1998).
\newblock Modeling failure time data by lehman alternatives.
\newblock {\em Communications in Statistics-Theory and methods},
  27(4):887--904.

\bibitem[G{\"u}rler, 1996]{gurler1996bivariate}
G{\"u}rler, {\"U}. (1996).
\newblock Bivariate estimation with right-truncated data.
\newblock {\em Journal of the American Statistical Association},
  91(435):1152--1165.

\bibitem[Hanagal, 2019]{hanagal2019shared}
Hanagal, D.~D. (2019).
\newblock Shared gamma frailty models based on reversed hazard.
\newblock In {\em Modeling Survival Data Using Frailty Models}, pages 191--211.
  Springer.

\bibitem[Hanagal, 2021]{hanagal2021correlated}
Hanagal, D.~D. (2021).
\newblock Correlated positive stable frailty models based on reversed hazard
  rate.
\newblock {\em Statistics in Biosciences}, pages 1--24.

\bibitem[Hanagal and Bhambure, 2017]{hanagal2017modeling}
Hanagal, D.~D. and Bhambure, S.~M. (2017).
\newblock Modeling australian twin data using shared positive stable frailty
  models based on reversed hazard rate.
\newblock {\em Communications in Statistics-Theory and Methods},
  46(8):3754--3771.

\bibitem[Keilson and Sumita, 1982]{keilson1982uniform}
Keilson, J. and Sumita, U. (1982).
\newblock Uniform stochastic ordering and related inequalities.
\newblock {\em Canadian Journal of Statistics}, 10(3):181--198.

\bibitem[Krishnamoorthy et~al., 2009]{krishnamoorthy2009model}
Krishnamoorthy, K., Mallick, A., and Mathew, T. (2009).
\newblock Model-based imputation approach for data analysis in the presence of
  non-detects.
\newblock {\em Annals of Occupational Hygiene}, 53(3):249--263.

\bibitem[Kundu and Gupta, 2010]{kundu2010class}
Kundu, D. and Gupta, R.~D. (2010).
\newblock A class of bivariate models with proportional reversed hazard
  marginals.
\newblock {\em Sankhya B}, 72(2):236--253.

\bibitem[Lawless, 2011]{lawless2011statistical}
Lawless, J.~F. (2011).
\newblock {\em Statistical models and methods for lifetime data}, volume 362.
\newblock John Wiley \& Sons.

\bibitem[Lehmann and Casella, 2006]{lehmann2006theory}
Lehmann, E.~L. and Casella, G. (2006).
\newblock {\em Theory of point estimation}.
\newblock Springer Science \& Business Media.

\bibitem[Paluszny and Gibson, 1974]{paluszny1974twin}
Paluszny, M. and Gibson, R. (1974).
\newblock Twin interactions in a normal nursery school.
\newblock {\em American Journal of Psychiatry}, 131(3):293--296.

\bibitem[Pandey et~al., 2020]{pandey2020analysis}
Pandey, A., Hanagal, D.~D., Gupta, P., and Tyagi, S. (2020).
\newblock Analysis of australian twin data using generalized inverse gaussian
  shared frailty models based on reversed hazard rate.
\newblock {\em International Journal of Statistics and Reliability
  Engineering}, 7(2):219--235.

\bibitem[Popovi{\'c} et~al., 2021]{popovic2021generalized}
Popovi{\'c}, B.~V., Gen{\c{c}}, A.~{\.I}., and Domma, F. (2021).
\newblock Generalized proportional reversed hazard rate distributions with
  application in medicine.
\newblock {\em Statistical Methods \& Applications}, pages 1--22.

\bibitem[Roy, 2002]{roy2002characterization}
Roy, D. (2002).
\newblock A characterization of model approach for generating bivariate life
  distributions using reversed hazard rates.
\newblock {\em Journal of the Japan Statistical Society}, 32(2):239--245.

\bibitem[Sankaran and Gleeja, 2006]{pg2006bivariate}
Sankaran, P. and Gleeja, V. (2006).
\newblock On bivariate reversed hazard rates.
\newblock {\em Journal of the Japan Statistical Society}, 36(2):213--224.

\bibitem[Wan, 2017]{wan2017simulating}
Wan, F. (2017).
\newblock Simulating survival data with predefined censoring rates for
  proportional hazards models.
\newblock {\em Statistics in medicine}, 36(5):838--854.

\bibitem[Ware and Demets, 1976]{ware1976reanalysis}
Ware, J.~H. and Demets, D.~L. (1976).
\newblock Reanalysis of some baboon descent data.
\newblock {\em Biometrics}, pages 459--463.

\end{thebibliography}


\begin{thebibliography}{}

\end{thebibliography}
\end{document}